\documentclass[aps,prl,superscriptaddress,twocolumn,showpacs]{revtex4}

\renewcommand*{\[}{\begin{equation}}

\renewcommand*{\]}{\end{equation}}

\def\PRA{{Phys.~Rev.~A} }

\def\JPB{{J.~Phys.~B} }
\def\PRL{{Phys.~Rev.~Lett.} }
\def\RMP{{Rev.~Mod.~Phys.} }
\def\JCP{{J.~Chem.~Phys.} }

\newcommand{\myscaleboxa}[1]{\scalebox{0.33}[0.33]{#1}}
\newcommand{\myscaleboxb}[1]{\scalebox{0.35}[0.35]{#1}}

\usepackage{epsfig}

\usepackage{hyperref}

\begin{document}

\title{Quantitative rescattering theory for laser-induced high-energy plateau photoelectron spectra}

\author{Zhangjin Chen}
\author{Anh-Thu Le}
\affiliation{J. R. Macdonald Laboratory, Physics Department, Kansas
State University, Manhattan, Kansas 66506-2604, USA}

\author{Toru Morishita}
\affiliation{Department of Applied Physics and Chemistry, University
of Electro-Communications, 1-5-1 Chofu-ga-oka, Chofu-shi, Tokyo,
182-8585, Japan and PRESTO, JST Agency, Kawaguchi, Saitama 332-0012,
Japan}

\author{C. D. Lin }
\affiliation{J. R. Macdonald Laboratory, Physics Department, Kansas
State University, Manhattan, Kansas 66506-2604, USA}

\date{\today}

\begin{abstract}

A comprehensive quantitative rescattering (QRS) theory for
describing the production of high-energy photoelectrons generated by
intense laser pulses is presented. According to the QRS, the
momentum distributions of these electrons can be expressed as the
product of a returning electron wave packet with the elastic
differential cross sections (DCS) between free electrons with the
target ion. We show that the returning electron wave packets are
determined mostly by the lasers only, and can be obtained from the
strong field approximation. The validity of the QRS model is
carefully examined by checking against accurate results from the
solution of the time-dependent Schr\"odinger equation for atomic
targets within the single active electron approximation. We further
show that experimental photoelectron spectra for a wide range of
laser intensity and wavelength can be explained by the QRS theory,
and that the DCS between electrons and target ions can be extracted
from experimental photoelectron spectra. By generalizing the QRS
theory to molecular targets, we discuss how few-cycle infrared
lasers offer a promising tool for dynamic chemical imaging with
temporal resolution of a few femtoseconds.

\end{abstract}


\maketitle

\section{I. Introduction}

Much of our knowledge of the nonlinear interaction of intense laser
radiation with atoms and molecules comes from the study of
above-threshold ionization (ATI) which is characterized by a
sequence of peaks in the electron spectrum, spaced by the photon
energy. Since its first observation \cite{AA4}, the subject has been
``reinvestigated'' many times. In 1987, as sub-picosecond laser
pulses became available, it was shown \cite{AA28} that ATI peaks
suffer significant energy shifts and broadening, and each peak
breaks up into substructures due to resonance enhancements produced
by ponderomotive shifts of states. These substructures are called
Freeman resonances. The nature of these Freeman resonances have been
carefully investigated, for example for Ar in Wiehle \emph{et al}.
\cite{AA26}. In recent years, with the introduction of COLTRIM
detectors where electrons are measured over almost the whole $4\pi$
angular region, the two-dimensional (2D) electron momentum spectra
or the longitudinal electron momentum spectra of the photoelectrons
or the target ions have been reported \cite{AA29b,AA29c}. These
measurements reveal considerable structure not only in the electron
energy distributions, but also in the angular distributions. All of
these studies focus on low energy electrons which are generated
either by multiphoton ionization mechanism or by the tunneling
ionization mechanism. According to the ``conventional'' wisdom,
depending on the Keldysh parameter, $\gamma=\sqrt{I_p/2U_p}$ where
$I_p$ is the ionization energy of the target and $U_p$ the
ponderomotive energy, if $\gamma$ is larger than one, the ATI
electrons are generated by multiphoton processes, while if $\gamma$
is small, tunneling ionization is responsible for producing the low
energy electrons. However, such a distinction is by no means
clear-cut. In Ref. \cite{AA29c}, the 2D electron momentum spectra
display pronounced fan-like structures even for laser intensities
well into the tunneling ionization regime. Theoretical studies of
electron momentum spectra \cite{chen_06,Arbo} obtained from solving
the time-dependent Schr\"{o}dinger equation (TDSE) show that even in
the tunneling region, photoelectron spectra show features that can
be identified with the absorption of integral number of photons, as
revealed by the angular distributions of the electron at fixed
energies. These theoretical multiphoton ionization features, when
convoluted with the effect of the spatial distribution of
intensities in a focused laser beam, can well reproduce the observed
experimental 2D electron momentum spectra \cite{toru_07}.

While low-energy electrons, with energy less than about $2U_p$,
account for the majority of the electrons generated by an intense
laser, already since 1993 photoelectrons extending to $10U_p$ or
more have also been observed. These electrons, unlike the low energy
electrons generated by the nonlinear processes, do not change with
the electron energy rapidly, until a new cutoff near about $10U_p$
is reached.  They are known as high-energy plateau photoelectrons.
Experiments showed that these electrons exhibit pronounced sidelobes
not seen in low-energy electrons
\cite{yang93,Gra03,Gaarde,Paulus_PRL94}. These high-energy ATI
(HATI) electrons have been interpreted as due to the rescattering
process \cite{Paulus_JPB94}. According to this model, electrons that
are freed from the target atom at some well-defined ionization time
may be driven back to revisit its parent ion. If these returning
electrons are backscattered by the target ion, they can be further
accelerated by the laser field and emerge as high-energy electrons,
reaching up to about $10U_p$. However, the plateau electron spectra,
with energies from $4U_p$ to about $10U_p$, are not always similar
for different targets. For  target like xenon, the plateau is flat,
but for other like krypton, the plateau drops steeply as the
electron energy increases. These features actually change with peak
laser intensities. Further studies of these HATI spectra around 1997
on inert gases discovered   that resonantlike enhancements occur in
the electron spectra for particular laser intensities. Depending on
the inert gas used, separate series of peaks have been observed.
These observations have generated a flurry of theoretical interest
\cite{milo-97}. Models based on analyzing results from solving TDSE
\cite{AA16,AA17} and from Floquet theory \cite{AA18,AA19} have been
proposed. Others are based on the channel-closing theory
\cite{Kopold,Tony}. More recent experiments confirmed that these
enhancement disappears when laser pulse duration is reduced
\cite{Gra03}. Since the HATI yields are four to five orders of
magnitude smaller, despite of these experimental investigations,
there are few systematic theoretical calculations in the literature.
In recent years, HATI electrons have drawn attention again since
when they are generated by few-cycle laser pulses, their counts on
the left and the right detectors along the laser polarization axis
are different. Such asymmetry can be used to determine the absolute
value of the carrier-envelope-phase (CEP) of the few-cycle pulses
\cite{paulus-cep}.

Recently, we investigated the 2D high-energy photoelectron momentum
spectra for atomic targets within the single active electron
approximation based on the well-known rescattering model
\cite{corkum,schaffer}. We proposed a quantitative rescattering
(QRS) theory \cite{toru08,chen_08} where the HATI electrons are
modeled as due to the backscattering of the returning electrons by
the target ion. According to the QRS, high-energy photoelectron
momentum distributions $D(k,\theta)$ is shown to be expressed simply
as
\begin{eqnarray}
\label{QRS}D(k,\theta)=W(k_r)\sigma(k_r,\theta_r)
\end{eqnarray}
where $\sigma(k_r,\theta_r)$ is the elastic differential cross
sections (DCS) between \emph{free} electrons, with momentum $k_r$,
with the target ion. Here $\theta_r$ is the scattering angle with
respect to the the direction of the returning electrons along the
laser polarization axis. In this equation, $W(k_r)$ is interpreted
as the momentum distribution of the returning electrons, to be
called returning wave packet (RWP) in this paper. The validity of
this QRS model has been tested using $D(k,\theta)$ calculated from
solving the TDSE, and $\sigma(k_r,\theta_r)$ from the standard
quantum mechanical scattering theory. Since there is a one-to-one
relation between $(k,\theta)$ and $(k_r,\theta_r)$, there are a
number of important results from the QRS, as reported in our recent
papers. It was shown in Morishita \emph{et al}.~\cite{toru08} and
Chen \emph{et al}.~\cite{chen_08} that one can extract elastic
scattering cross sections $\sigma(k_r,\theta_r)$ between free
electrons and atomic ions from the HATI electron momentum spectra.
The predictions have been confirmed in three recent experiments
\cite{ueda,ray,sam-prl}. In Chen \emph{et al}. \cite{chen_07}, it
was further shown that the momentum distribution $W(k_r)$ of the RWP
can be extracted from the second-order strong field approximation
(SFA2), and that $W(k_r)$ depends very little on the target (i.e.,
up to an overall normalization which is related to the total
ionization probability). Thus one can use SFA2 to obtain $W(k_r)$.
By multiplying it with $\sigma(k_r,\theta_r)$, we can use
Eq.~(\ref{QRS}) to obtain accurate high-energy photoelectron
momentum distribution $D(k,\theta)$. Since $W(k_r)$ depends mostly
on the lasers only, the target dependence of the HATI spectra can
thus be explained based on the behavior of the elastic scattering
cross sections $\sigma(k_r,\theta_r)$. Based on the QRS model, the
energy dependence of plateau ATI electrons seen for different
targets are easily understood \cite{chen_08}. For a given target but
different lasers, the HATI momentum spectra are determined by the
RWP. Applying the QRS model to few-cycle pulses, where the RWP
varies with the change of the CEP, we have shown that the QRS theory
can be easily used to retrieve the absolute value of the CEP
\cite{sam-prl,sam-jpb}. Since nonsequential double ionization of
atoms and molecules are understood based on the rescattering
mechanism, in Micheau \emph{et al}. \cite{sam-NSDI} we showed that
using the wave packet $W(k_r)$ obtained from the HATI spectra, we
can use QRS to obtain nonsequential double ionization yields.

In this paper, we provide the full details of the QRS theory on
atomic targets within the single active electron approximation and
establish its validity. Clearly our goal is not to limit ourselves
to atomic targets only. We would like to study HATI spectra from
molecular targets as well, in particular, from transient molecules.
Recall that HATI electrons result from backscattering of the
returning electrons by the target ion, i.e., electrons that undergo
hard collisions with the target. Thus one should be able to retrieve
the structure information of the target from the HATI spectra. In
fact, we have shown that this is indeed possible for atomic targets
already \cite{XU09}. Since laser pulses of duration of a few
femtoseconds are readily available, one can perform pump-probe
measurements where the pump pulse  initiates a chemical reaction,
such that the atomic coordinates of the molecule would evolve in
time. Using a probe laser to take HATI spectra at different time
delays, one will then have the opportunity to extract the structure
of the transient molecule as a function of time from the measured
HATI spectra. Thus short laser pulses may serve as a powerful tool
for dynamic chemical imaging of transient molecules, with temporal
resolution of a few femtoseconds.

The rest of this paper is arranged as follows. In the next section,
we discuss how to calculate the HATI electron momentum spectra by
solving the TDSE and using the SFA2. We also explain how the elastic
differential cross sections are calculated. We then establish the
QRS model. The validity of the QRS model is carefully examined in
Section III, by testing against results obtained from solving the
TDSE. In Section IV we illustrate the application of the QRS model
to experimental HATI spectra. To compare with experimental electron
energy spectra, we include the laser focus volume effect. We finish
the paper with a summary and outlook. We point out that a similar
QRS model has been developed for high-order harmonic generation
(HHG) \cite{toru08,atle08,atlejpb}.

Atomic units are used in this paper unless otherwise noted. We also
mention that in all the calculations the CEP is set to zero in this
paper.

\section{II. Theoretical background}

The theory part is separated into four sections. Since the concept
of rescattering can be understood in classical mechanics, we first
consider the classical rescattering theory for an electron in a
one-dimensional (1D) monochromatic laser field. Then we discuss the
calculations of ATI spectra by solving the TDSE and using the SFA2.
For completeness we also include how the  elastic scattering cross
sections are computed.

\subsection{A. Classical one-dimensional rescattering theory}

A classical 1D rescattering theory has been discussed by Paulus
\emph{et al}.~\cite{Paulus_JPB94}. Suppose that an electron in the
1D atom is first released at some time $t_0$ into a monochromatic
laser field $\textbf{E}(t)=\hat{z}E_0\cos \omega t$, the Newton's
equation of motion for this system is given by
\begin{eqnarray}
\ddot{z}(t)=-E_0\cos\omega t,
\end{eqnarray}
consequently the position of the electron at time $t$ reads
\begin{eqnarray}
z(t)&=&z(t_0)+\frac{E_0}{\omega^2}\left[\cos \omega t-\cos\omega
t_0\right]\nonumber \\
&&+\left[\frac{E_0}{\omega}\sin\omega
t_0+\dot{z}(t_0)\right](t-t_0).
\end{eqnarray}
If the electron is initially at the origin with zero initial
velocity, the time $t_r$ at which it returns to the origin satisfies
\begin{eqnarray}
\label{t_r}\cos\omega t_r-\cos\omega t_0+ \omega\sin\omega
t_0(t_r-t_0)=0.
\end{eqnarray}
The electron will never return to the origin if it is ionized before
the laser field reaches its peak value while it can return to the
origin more than once when it is born after the peak. It has been
shown that higher order returns make very small contribution to the
yield \cite{chen_07}. We consider only the first return here. The
momentum $k_r$ of the electron when it first returns to the origin
at time $t_r$ is
\begin{eqnarray}
k_r\equiv \dot{z}(t_r)=-\frac{E_0}{\omega}\left(\sin\omega
t_r+\sin\omega t_0\right).
\end{eqnarray}

\begin{figure}
\mbox{\rotatebox{270}{\myscaleboxa{
\includegraphics{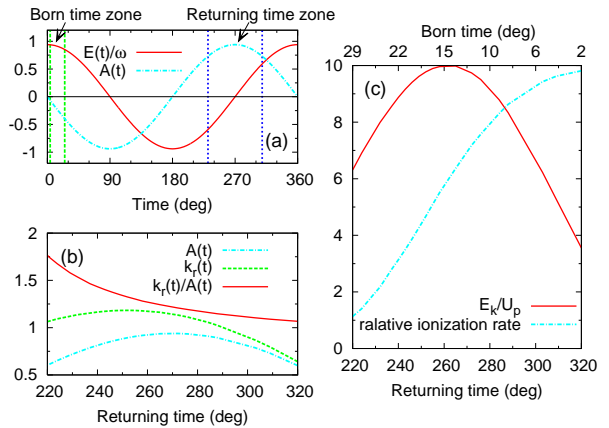}}}}
\caption{(Color online) Classical model of a 1D  electron in a
monochromatic laser field. (a) Electric field and vector potential
for a laser at the intensity of $1.0\times10^{14}$ W/cm$^2$ and
wavelength of 800~nm. The born time and returning time zones are
marked. (b) The electron velocity $k_r$, vector potential $A$, and
their ratio at the time of return t=t$_r$, within the returning time
zone. (c) Photoelectron energy after it has been backscattered by an
angle $\theta_r=180^\circ$ against the returning time (bottom
horizontal axis). Also shown is the relative ionization rate for
electrons released with respect to the born time (top horizontal
axis).}
\end{figure}

In Fig.~1(a) the electric field and the vector potential are plotted
vs $\omega t$ for a typical set of laser parameters, say
$I_0=1.0\times10^{14}$ W/cm$^2$ and $\lambda=800$ nm. The maximum
value of the vector potential is $A_0=0.94$. In the rescattering
model, consider electrons that return within $231^{\circ}<\omega
t_r<309^{\circ}$. In this region, $A(t_r)>0.78A_0$ and by solving
Eq.~(3), the corresponding born time is within $4^{\circ}<\omega
t_0<25^{\circ}$. Both the born time zone and returning time zone are
marked in Fig.~1(a). The electron born at time $\omega
t_0=13^{\circ}$ returns at time $\omega t_r=270^{\circ}$. It should
be noted that electrons born before $13^{\circ}$ return at the time
after $270^{\circ}$ and they follow a long trajectory, while those
born after $13^{\circ}$ return before $270^{\circ}$ and they follow
a short trajectory.

The vector potential, the velocity of the electron at the return
time, and the ratio of the returning velocity to the vector
potential are plotted in Fig.~1(b), in the returning time zone. It
can be seen that $k_r(t)/A(t)$ decrease from 1.76 at $220^{\circ}$
to 1.07 at $320^{\circ}$.

We next assume that the electron undergoes collision with the target
ion and elastically scattered by an angle $\theta_r$ with respect to
its incident direction. For $t\geq t_r$, the components of its
velocity, along the polarization axis and perpendicular to it, are
given by
\begin{eqnarray}
\dot{z}(t)&=&-\frac{E_0}{\omega}\left[\sin\omega t-\sin\omega
t_1+\cos\theta_r\left(\sin \omega t_r\sin \omega t_0\right) \right],
\nonumber \\
\dot{y}(t)&=&-\frac{E_0}{\omega} \left(\sin\omega t_r-\sin\omega
t_0\right).
\end{eqnarray}
From the above equation, the photoelectron energy $E_k$ measured by
the detector outside of the field can be obtained by subtracting the
ponderomotive potential, $U_p=E_0^2/(4\omega^2)$, from the time
averaged kinetic energy
\begin{eqnarray}
E_k=2U_p[\sin^2\omega t_0+2\sin\omega t_r(1-\cos\theta_r)(\sin\omega
t_1 - \sin \omega t_0)].\label{Energy}
\end{eqnarray}

It is easy to find from (\ref{Energy}) that, if the electron is born
at $\omega t_0=14^{\circ}$, it returns to the origin at $\omega
t_r=265^{\circ}$ when the vector potential almost reaches the
maximum, then the photoelectron will have the maximal energy
$E_k^{\text {max}}=10.007 U_p$ provided that the returning electron
experiences a backwardscattering  of $\theta_r=180^\circ$. The
electron's kinetic energy $E_k$ measured by the detector along the
polarization axis is shown in in Fig.~1(c), together with the
returning time versus with the relative ionization rate of electrons
released in the born time zone. The lower horizontal axis in
Fig.~1(c) indicates the returning time while the corresponding born
time is given on the top. In the QRS model, we investigate
backscattered electrons with energies greater than $4U_p$. From
Fig.~1(c), it can be seen that electrons return at two different
times could have the same kinetic energy except for the time around
$265^{\circ}$ at which $E_k$ has the maximal value. Actually,
electrons which return after $265^{\circ}$ are born before
$14^{\circ}$. These electrons have higher ionization rate than those
born after $14^{\circ}$. For example, the ionization rate for
electrons which are born at $22^{\circ}$ and return around
$240^{\circ}$ is about 3 times smaller than   electrons born at
$2^{\circ}$ and return at $320^{\circ}$.

Let us look back at Fig.~1(b) which shows that the ratio
$k_r(t)/A(t)$  becomes flatter after a sharp decrease until
$240^{\circ}$ and the mean value of $k_r(t)/A(t)$ in the returning
time range from $240^{\circ}$ to $320^{\circ}$ is about 1.25.
Although the ratio $k_r(t)/A(t)$ for returning time less than
$240^{\circ}$  deviates more from this mean value, the electron
yield from this part is very small, as seen in Fig.~1(c).

\subsection{B. Method of solving the time-dependent Schr\"{o}dinger equation}

The method for solving the time-dependent Schr\"{o}dinger equation
has been described in our previous works \cite{chen_06,toru_07}
where we studied the low-energy electron momentum spectra. Much more
effort is needed to obtain accurate momentum spectra for high energy
electrons. Here we describe the essential steps of the calculations.

We treat the target atom in the single active electron model. The
Hamiltonian for such an atom in the presence of a linearly polarized
laser can be written as
\begin{eqnarray}
\label{tdse2} H=H_0+H_i(t)=-\frac{1}{2}\nabla^2 +V(r)+H_i(t).
\end{eqnarray}
The atomic model potential $V(r)$ is parameterized in the form
\begin{eqnarray}
\label{potential0}V(r)=-\frac{1+a_{1}e^{-a_{2}r}+a_{3}re^{-a_{4}r}+a_{5}e^{-a_{6}r}}{r}.
\end{eqnarray}
The parameters in Eq.~(\ref{potential0}) are obtained by fitting the
calculated binding energies from this potential to the experimental
binding energies of the ground state and the first few excited
states of the target atom. The parameters for the targets used in
this paper can be found in \cite{tong05}. For Kr and Xe, we use the
potential given by Garvey \emph{et al}. \cite{Green}. The model
potential for a neutral atom can also be expressed as
\begin{eqnarray}
\label{potential1}V(r)= V_{\text{s}}(r)-1/r,
\end{eqnarray}
where $V_{\text{s}}(r)$ is a short-range potential. The atom-field
interaction $H_i(t)$, in length gauge, is given by
\begin{eqnarray}
\label{tdse3}H_i=\textbf{r}\cdot\textbf{E}(t).
\end{eqnarray}
For a linearly polarized laser pulse (along the $z$ axis) with
carrier frequency $\omega$ and the CEP, $\varphi$, the field is
taken to have the form
\begin{eqnarray}
\textbf{E}(t)=\hat{z}E_{0}\cos^{2}\left(\frac{\pi
t}{\tau}\right)\cos(\omega t+\varphi)
\end{eqnarray}
for the time interval ($-\tau$/2, $\tau$/2) and zero elsewhere. The
pulse duration, defined as the full width at half maximum (FWHM) of
the intensity, is given by $\Gamma=\tau/2.75$.

The time evolution of the electronic wavefunction
$\Psi(\textbf{r},t)$, which satisfies the TDSE,
\begin{eqnarray}
\label{tdse1}i\frac{\partial}{\partial
t}\Psi(\textbf{r},t)=H\Psi(\textbf{r},t)
\end{eqnarray}
is solved by expanding in terms of eigenfunctions,
$R_{nl}(r)Y_{lm}(\mathbf{\hat{r}})$, of $H_0$, within the box of
$r\in[0,r_{\text{max}}]$
\begin{eqnarray}
\label{expansion}\Psi(\textbf{r},t)=\sum_{nl}c_{nl}(t)R_{nl}(r)Y_{lm}(\mathbf{\hat{r}})
\end{eqnarray}
where the radial functions $R_{nl}(r)$ are expanded by the DVR
(discrete variable representation) \cite{DVR1,DVR2,DVR3} basis set
associated with Legendre polynomials, while the $c_{nl}$ are calculated
using the split-operator method~\cite{tong97b}
\begin{eqnarray}
\label{time}c_{nl}(t+\Delta t)&\simeq& \sum_{n'l'}\{e^{-iH_0 \Delta
t/2}e^{-iH_i(t+\Delta t/2)\Delta t} \nonumber \\
&&\times e^{-iH_0 \Delta t/2}\}_{nl,n'l'} c_{n'l'}(t)
\end{eqnarray}
where the matrix elements are evaluated efficiently by using the DVR
quadrature. For short pulses, say, $\Gamma=8$~fs, and for electron
energy as high as $12U_p$, converged results can be obtained by
setting $r_{\text{max}}=1200$. Note that in (\ref{expansion}), only
$m=0$ is taken into account since for linearly polarized laser
pulses, contribution to the ionization probability from $m=\pm1$ is
relatively much smaller in comparison to the $m=0$ component.

The photoelectron yield is computed at the end of the laser pulse by
projecting the total final wave function onto eigenstates of a
continuum electron with momentum $\textbf{k}$,
\begin{eqnarray}
\label{momentum}D(k,\theta)\equiv\frac{\partial^3
P}{\partial^3\textbf{k}}=|\langle
\Phi_{\textbf{k}}^{-}|\Psi(t=\tau/2)\rangle|^2
\end{eqnarray}
where the continuum state $\Phi_{\textbf{k}}^{-}$ satisfies the
following equation
\begin{eqnarray}
\label{phip}\left[-\frac{1}{2}\nabla^2
+V(r)\right]\Phi_{\textbf{k}}^{-}=\frac{k^2}{2}\Phi_{\textbf{k}}^{-}.
\end{eqnarray}

\subsection{C. Strong field approximation for calculating ATI electron spectra}

While direct solution of the Schr\"{o}dinger equation in a
time-dependent laser field has been widely used, the simpler strong
field approximation is of interest for analyzing features of intense
laser-atom interactions.

By treating electron-laser interaction as the strong field and
electron-target ion interaction as a perturbation, the amplitude for
generating a photoelectron with momentum $\textbf{k}$ is given by
\begin{eqnarray}
\label{full_SFA}f(\textbf{k})=f_1(\textbf{k})+f_2(\textbf{k})
\end{eqnarray}
where we included the first two terms of the perturbation series
only.

In Eq.~(\ref{full_SFA}), the first order term $f_1(\textbf{k})$,
which is traditional called the strong field approximation (SFA),
will be called SFA1, to be distinguished from the second term
$f_2(\textbf{k})$, which will be called SFA2. The SFA1 is given by
\begin{eqnarray}
\label{1st-order0}f_1(\textbf{k})=-i
\int_{-\infty}^{\infty}dt\left\langle\chi_{\textbf{k}}(t)\left|H_i(t)\right|\Psi_{0}(t)\right\rangle
\end{eqnarray}
where $\Psi_{0}$ is the ground state wavefunction. The Volkov state $\chi_{\textbf{k}}$ is given by
\begin{eqnarray}
\left\langle\textbf{r}|\chi_{\textbf{k}}(t)\right\rangle=\frac{1}{(2
\pi)^{3/2}}e^{i\left [\textbf{k}+\textbf{A}(t)\right ] \cdot
\textbf{r}} e^{-iS({\textbf{k}},t)}
\end{eqnarray}
where the action S is
\begin{eqnarray}
S(\textbf{k},t)=\frac{1}{2}\int_{-\infty}^t
dt^{\prime}\left[\textbf{k}+\textbf{A}(t^{\prime})\right]^2.
\end{eqnarray}
The SFA2 term in (\ref{full_SFA}) is expressed as
\begin{eqnarray}
\label{2nd-order0}f_2(\textbf{k})&=&-\int_{-\infty}^{\infty}dt\int_{t}^{\infty}dt^{\prime}\int
d\textbf{p}\left\langle\chi_{\textbf{k}}(t^{\prime})\left|V\right|\chi_{\textbf{p}}(t^{\prime})\right\rangle \nonumber \\ && \times
\left\langle\chi_{\textbf{p}}(t)\left|H_i(t)\right|\Psi_{0}(t)\right\rangle.
\end{eqnarray}
It consists of three time-ordered steps by the electron: tunnel
ionization, propagation in the laser field, and elastic scattering
with the parent ion. Note that the SFA2 used here is identical to
the so-called improved strong field approximation
\cite{milo-jpb,has-07}.

To evaluate the SFA1 amplitude, we rewrite Eq.~(\ref{1st-order0}) as
\begin{eqnarray}
\label{1st-order}f_1(\textbf{k})&=&-i\frac{1}{(2\pi)^{3/2}}
\int_{-\infty}^{\infty}dtE(t)e^{iS(\textbf{k},t)} e^{iI_pt}
\nonumber \\
&\times&\int d\textbf{r}e^{-i[\textbf{k}+\textbf{A}(t)]\cdot
\textbf{r}} r\cos \theta \Psi_{0}(\textbf{r})
\end{eqnarray}
where $I_p$ is the ionization potential of the ground state
$\Psi_{0}(\textbf{r})$, and $\theta$ is the polar angle. The ground
state wavefunction is calculated from the model potential $V(r)$. To
perform  integration over space coordinates in (\ref{1st-order}), we
use the identity
\begin{equation}
e^{-i\textbf{q}\cdot\textbf{r}}=4\pi\sum_{lm}
i^{-l}j_l(qr)Y_{lm}(\hat{\textbf{r}})Y_{lm}^{\ast}(\hat{\textbf{q}})
\label{plane}
\end{equation}
where $j_l(qr)$ is the spherical Bessel function and
$\cos\theta=\sqrt{4\pi/3}Y_{10}(\hat{\textbf{r}})$. Consequently,
the integral over space coordinates can be expressed
\begin{eqnarray}
\Psi_{0}(\textbf{q})&\equiv&\int d\textbf{r}e^{-i\textbf{q}\cdot
\textbf{r}}
r\cos \theta \Psi_{0}(\textbf{r}) \nonumber \\
&=&4\pi\sqrt{\frac{4\pi}{3}}\sum_{lm}i^{-l}Y_{lm}(\hat{\textbf{q}})\int
dr
r^3 R_{n_0l_0}(r)j_l(qr)\nonumber \\
&\times&\int
d\hat{\textbf{r}}Y_{lm}^{\ast}(\hat{\textbf{r}})Y_{10}(\hat{\textbf{r}})Y_{l_0m_0}(\hat{\textbf{r}})
\end{eqnarray}
where the initial state wave function $\Psi_{0}(\textbf{r})=R_{n_0l_0}(r)Y_{l_0m_0}(\hat{\textbf{r}})$. Due to reason mentioned before, for
linearly polarized laser field, we consider $m_0=0$ only, and
\begin{eqnarray}
&&\int
d\hat{\textbf{r}}Y_{lm}^{\ast}(\hat{\textbf{r}})Y_{10}(\hat{\textbf{r}})Y_{l_0m_0}(\hat{\textbf{r}})
\nonumber \\
&&=\sqrt{\frac{3(2l_0+1)}{4\pi(2l+1)}}C(1l_0l;000)C(1l_0l;000)\delta_{m0}
\end{eqnarray}
where the C's are the Clebsch-Gordan coefficients. The remaining
integration over $r$ is done analytically if hydrogenic wavefunction
is used, otherwise it is evaluated numerically. The integration over
time is carried out numerically.

For the SFA2 amplitude, we used saddle point approximation for the
integral over the momentum $\textbf{p}$ of the intermediate states
and  Eq.~(\ref{2nd-order0}) becomes
\begin{eqnarray}
\label{2nd-order}f_2(\textbf{k})&=&-\int_{-\infty}^{\infty}dt\int_{-\infty}^{t}dt^{\prime}
\left[\frac{2\pi}{\epsilon+i(t-t^{\prime})}\right]^{3/2}E(t^{\prime})e^{iI_p t^{\prime}} \nonumber \\
&\times&e^{-i[S(\textbf{p}_s,t)-S(\textbf{k},t)]}e^{iS(\textbf{p}_s,t^{\prime})} \nonumber \\
&\times& \frac{1}{(2\pi)^3}\int
d\textbf{r}^{\prime}e^{i(\textbf{p}_s-\textbf{k})\cdot
\textbf{r}^{\prime}} V\left(\textbf{r}^{\prime}\right)\nonumber \\
&\times&\frac{1}{(2\pi)^{3/2}}\int
d\textbf{r}e^{-i\left[\textbf{p}_s+\textbf{A}(t^{\prime})\right]\cdot\textbf{r}}
r \cos\theta \Psi_0(\textbf{r}).
\end{eqnarray}
The saddle point is calculated with respect to quasiclassical
actions only
\begin{eqnarray}
\label{sad_k}\textbf{p}_s(t,t^{\prime})=-\frac{1}{t-t^{\prime}}
\int_{t^{\prime}} ^t dt^{\prime\prime}\textbf{A}(t^{\prime\prime}),
\end{eqnarray}
and the related actions are given by
\begin{eqnarray}
\label{action1}S(\textbf{p}_s,t)=\frac{1}{2}\int_{-\infty}^t
dt^{\prime\prime}\left[\textbf{p}_s(t,t^{\prime})+\textbf{A}(t^{\prime\prime})\right]^2,
\end{eqnarray}
and
\begin{eqnarray}
\label{action2}S(\textbf{p}_s,t^{\prime})=\frac{1}{2}\int_{-\infty}^{t^{\prime}}
dt^{\prime\prime}\left[\textbf{p}_s(t,t^{\prime})+\textbf{A}(t^{\prime\prime})\right]^2.
\end{eqnarray}

The arbitrary small parameter $\epsilon$ in (\ref{2nd-order}) is
introduced to remove possible singularity when $t\rightarrow
t^{\prime}$. Actually, the integral (\ref{2nd-order}) converges for
the case of initial state $\Psi_0(\textbf{r})$ having $S$ symmetry
(as for H) while it is divergent for the case of $P$ symmetry (as
for Ar) without $\epsilon$ \cite{zhou_08}.   The Fourier transform
of the potential $V(r)$ in (9) is given by
\begin{eqnarray}
\label{pot_Fourier}&&V(\textbf{q})\equiv\int d\textbf{r}
\exp(i\textbf{q}\cdot
\textbf{r}) V(r)\nonumber \\
&&=-4\pi\left[\frac{1}{q^2}+\frac{a_1}{a_2^2+q^2}+\frac{2a_3a_4}{(a_4^2+q^2)^2}+\frac{a_5}{a_6^2+q^2}\right]
\end{eqnarray}
It is obvious from (\ref{pot_Fourier}) that the Fourier transform of
$V(r)$ diverges when $q\rightarrow0$. Therefore, in  actual
numerical calculations, we multiply the potential by a damping
factor
\begin{eqnarray}
\label{short_pot}\tilde{V}(r)=V(r)e^{-\alpha r}
\end{eqnarray}
to avoid the singularity of the integral. We checked that the
results mainly affect  the magnitude but not the shape of the HATI
spectra.

\subsection{D. Elastic differential elastic cross sections}

In this section, we briefly summarize the standard potential
scattering theory which has been well documented in the textbooks,
see Ref. \cite{schiff,joachain}, for example. Without loss of
generality, here we address elastic scattering of an electron by a
spherical potential $V(r)$  by solving the time-independent
Schr\"{o}dinger equation
\begin{eqnarray}
\label{shr}[\nabla^2+k^2-U(r)]\psi(\textbf{r})=0
\end{eqnarray}
where $U(r)=2V(r)$ is the reduced potential and $k$ is the electron
momentum, related to the incident electron energy by
$k=\sqrt{2E}$. For   short-range potential which tends to zero
faster then $r^{-2}$ as $r\rightarrow\infty$, the scattering wave
function satisfies the asymptotic outgoing wave boundary condition
\begin{eqnarray}
\psi^{+}(\textbf{r})_{r\rightarrow\infty}=
\frac{1}{(2\pi)^{3/2}}\left[\exp(ikz)+f(\theta)\frac{\exp(ikr)}{r}\right]
\label{boundary}
\end{eqnarray}
where $\theta$ is the polar angle measured from the incident
direction. We choose the $z$-axis along the direction of the incident wave
vector $\textbf{k}$.

We solve (\ref{shr}) by expanding the scattering wave function into partial waves,
\begin{equation}
\psi^{+}(\textbf{r})=\sqrt{\frac{2}{\pi}}\frac{1}{kr}\sum_{lm} i^{l} e^{i\delta_l} u_l(kr)Y_{lm}(\hat{\textbf{r}})Y_{lm}^{\ast}(\hat{\textbf{k}})
\label{distort1n}
\end{equation}
where $Y_{lm}$ is a spherical harmonic. The continuum waves are
normalized to $\delta(\textbf{k}-\textbf{k}^{\prime })$. The radial
function $u_l(kr)$ satisfies
\begin{eqnarray}
\label{shr_r1}\left[\frac{d^2}{dr^2}+k^2-\frac{l(l+1)}{r^2}-U(r)\right]u_l(kr)=0.
\end{eqnarray}
For a plane wave, when $U(r)=0$, the radial component $u_l(kr)/kr$
in (\ref{distort1n}) is the standard spherical Bessel function
$j_{l}(kr)$.

The radial part of the scattering wave, $u_l(k,r)$, has the
asymptotic form
\begin{eqnarray}
\label{ubound1} u_l(kr)\rightarrow\sin(k r-\frac{1}{2}l\pi+\delta_l)
\end{eqnarray}
where the phase shift $\delta_l$ reflects the influence of the interaction.

The above equations are valid for short-range potentials only. For
an electron in a Coulomb potential $V_{\text c}=-Z/r$, its full
wavefunction can be expanded as
\begin{eqnarray}
\psi_{\text{c}}^{+}(\textbf{r})=\sqrt{\frac{2}{\pi}}\frac{1}{kr}\sum_{lm}
i^{l}e^{i\sigma_l}u^{\text{c}}_l(kr)Y_{lm}(\hat{\textbf{r}})Y_{lm}^{\ast}(\hat{\textbf{k}})
\label{distort2}
\end{eqnarray}
where
\begin{eqnarray}
\sigma_l=\arg[\Gamma(l+1+i\eta)]
\end{eqnarray}
is called the Coulomb phase shift with $\eta=-Z/k$. The radial
wavefunction $u^{\text{c}}_l(kr)$ is solved from
\begin{eqnarray}
\label{shr_r2}\left[\frac{d^2}{dr^2}+k^2-\frac{l(l+1)}{r^2}-\frac{2\eta
k}{r}\right]u^{\text{c}}_l(kr)=0.
\end{eqnarray}
However, the expansion (\ref{distort2}) does not converge well for a
long-range Coulomb potential. For pure Coulomb scattering, the
treatment in parabolic coordinates is simpler and the scattering
amplitude is given by
\begin{eqnarray}
\label{pure_C}f_{\text{c}}(\theta)=-\eta
\exp(2i\sigma_0)\frac{\exp\{-i\eta
\ln[\sin^2(\theta/2)]\}}{2k\sin^2(\theta/2)}.
\end{eqnarray}

For electron-atomic target ion scattering within the single active
electron model, the model potential is written as the sum of the
Coulomb potential with $Z=1$ and a short-range potential $V_{\text
s}(r)$, see Eq.~(\ref{potential1}). For such a modified Coulomb
potential problem, the scattering amplitude is given by
\begin{eqnarray}
\label{amp}f(\theta)=f_{\text c}(\theta)+\hat{f}(\theta)
\end{eqnarray}
where the first term is the scattering amplitude by the Coulomb
potential alone [Eq.~(\ref{pure_C})], and the second term is given
by
\begin{eqnarray}
\label{amp22}\hat{f}(\theta)=\sum_{l=0}^{\infty} \frac{2l+1}{k}
\exp(2i\sigma_l)\exp(i\delta_l)\sin\delta_l P_l(\cos\theta)
\end{eqnarray}
where the $P_l(\cos\theta)$ are Legendre polynomials, and $\delta_l$
is the phase shift from the short-range potential. Due to the short
range nature, the summation in Eq.~(\ref{amp22}) can be truncated
after some number of partial waves, depending the electron energy.
The elastic scattering DCS is then given by
\begin{eqnarray}
\label{dcs}\sigma(k,\theta)\equiv\frac{dP}{d\Omega}=|f_{\text
c}(\theta)+\hat{f}(\theta)|^2.
\end{eqnarray}

For high-energy collisions, one may calculate the differential cross
sections using the first Born approximation, or the plane wave Born
approximation (PWBA), in which, the DCS is given by
\begin{eqnarray}
\label{pwba}
\sigma_{\text{PWBA}}(k,\theta)=\frac{1}{4\pi^2}|V(\textbf{q})|^2
\end{eqnarray}
where $\textbf{q}$ is the momentum transfer and its magnitude is
$q=2k\sin(\theta/2)$. In PWBA the continuum electron wavefunctions
are represented by  plane waves. For electron-target ion collisions,
PWBA is not valid even at large collision energies since it neglects
the effect of long-range Coulomb interaction as well as the strong
short-range potential due to the atomic ion or molecular ion core.
In SFA2,  scattering of the returning electron wave packet by the
target ion is treated by the plane wave Born approximation. This
limits the accuracy of using SFA2 in describing the HATI spectra.

\section{III. Quantitative rescattering theory and its region of validity}

\subsection{A. Features of ATI electron energy and momentum spectra}

In the ATI photoelectron energy spectra, it has been well recognized
that, after a sharp decrease to around $3U_p$, a plateau exists from
4 to $10U_p$. This universal phenomenon has been observed in
experiment, and in the TDSE and SFA calculations as well, see
Fig.~2. These figures also show that above about $4U_p$, SFA2
dominates the total electron spectra. Since SFA2 contains a
first-order interaction between the electron and the target ion [see
Eq.~(\ref{2nd-order0})], it is appropriate to attribute that ATI
electrons above $4U_p$ are produced by the rescattering processes.
In Fig.~2 we note that the HATI spectra for Xe target obtained from
TDSE are quite different from H when they are exposed to the same
laser pulse. The HATI plateau in Xe is very flat. It remains almost
constant between 4.5-$10U_p$, while for H target, in the same energy
region the yield drops by a large factor.

\begin{figure}
\mbox{\rotatebox{270}{\myscaleboxa{
\includegraphics{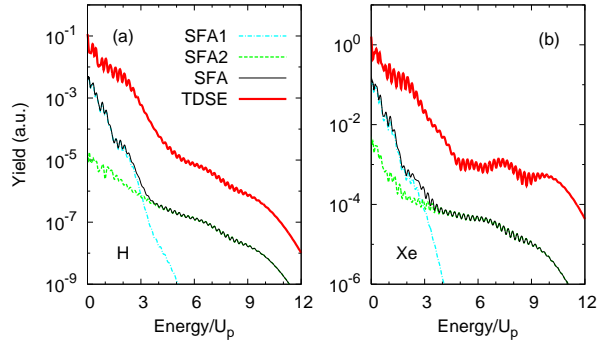}}}}
\caption{(Color online) Angle-integrated energy spectra (in units of
the ponderomotive energy $U_p$) calculated from SFA (SFA1 and SFA2)
compared with those by solving the TDSE for single ionization of (a)
H and (b) Xe in a 5~fs laser pulse at the intensity of
$1.0\times10^{14}$ W/cm$^2$ with the wavelength of 800~nm.}
\end{figure}

Since ATI electrons are produced mostly along the direction of
laser polarization (we consider linear polarization only), at
photoelectron energies where rescattering becomes dominant depend on
the angle of the photoelectrons. In Fig.~3, we show the electron
energy distributions for electrons emitted at 0, 30, 60 and 90
degrees with respect to   laser polarization, calculated using
SFA1 and SFA2, respectively. We note that the cutoff for SFA1 where
SFA2 becomes dominant shifts from about 4$U_p$ at zero degree to
3$U_p$, 2$U_p$, 1.5$U_p$, respectively, at the angles of 30, 60 and
90 degrees. The laser parameters used in the calculations for
Figs.~2 and 3 are given in the captions.

\begin{figure}
\mbox{\rotatebox{270}{\myscaleboxa{
\includegraphics{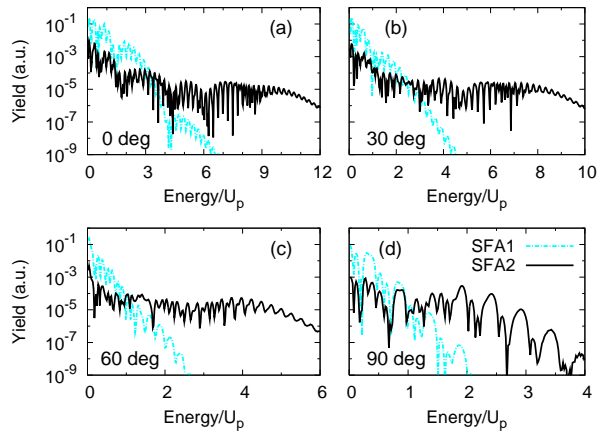}}}}
\caption{(Color online) SFA1 and SFA2 energy spectra for single
ionization of Xe in a 10~fs laser pulse at the intensity of
$1.0\times10^{14}$ W/cm$^2$ with the wavelength of 800~nm at angles
of (a) 0$^\circ$, (b) 30$^\circ$, (c) 60$^\circ$, and (d)
90$^\circ$. The SFA2 dominant region moves to lower energies as the
angle is increased.}
\end{figure}

Another method of presenting the angular dependence of electron
energy distribution is to display the 2D momentum distributions. In
Fig.~4, the 2D electron momentum distributions obtained from TDSE
and SFA2 are shown using the same laser parameters of Fig.~2. Only
the large momentum portion is considered since the inside is
dominated by SFA1, i.e., the direct ionization. First we note that
the appearance of circular ``bands'' at large momenta for the H
target, in both calculations. The centers of these semi-circular
rings are not at the origin, but are shifted along the polarization
axis, one on each side. The rings are very similar for H and Xe in
SFA2, but in TDSE, the intensity distributions show clear structure
in Xe, in particular, clear minima at some angles.

\begin{figure}
\mbox{\rotatebox{270}{\myscaleboxa{
\includegraphics{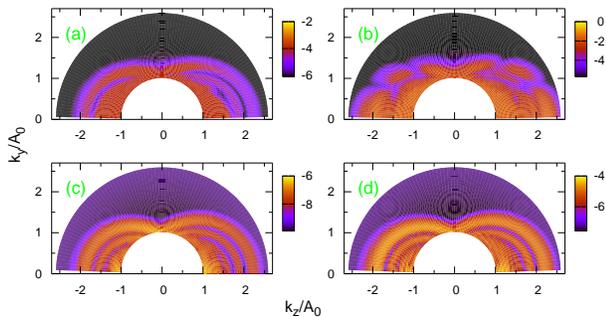}}}}
\caption{(Color online) 2D electron momentum distributions for
single ionization of H and Xe in a 5~fs laser pulse at the intensity
of $1.0\times10^{14}$ W/cm$^2$ with the wavelength of 800~nm. (a)
TDSE results for H; (b) TDSE results for Xe; (c) SFA2 results for H;
(d) SFA2 results for Xe.}
\end{figure}

\subsection{B. Extracting electron-target ion elastic differential cross sections from HATI spectra}

High-energy ATI electrons have been observed since 1993
\cite{yang93,Paulus_PRL94,Paulus_JPB94,Gra03,Gaarde}. They were
interpreted as due to the backscattering of the returning electrons
by the target ion. Indeed classical simulation \cite{Paulus_JPB94}
shows that electrons which return with maximum kinetic energy of
$3.17U_p$, if backscattered by 180 degrees, would emerge with
kinetic energy of about $10U_p$. Let $A_0$ be the peak value of the
vector potential of the laser pulse, $U_p=A_0^2/4$. For an electron
that returns at $3.17U_p$, it has momentum $k_r=1.26A_0$. For a beam
of electrons with momentum $k_r$, after elastically scattered, the
momentum space forms a circle in 2D (or a surface in 3D) of radius
$k_r$. Since scattering occurs when the laser field is nearly zero
and the vector potential almost has the maximum value $A_0$, each
electron will gain an additional drift momentum $A_0$ as it emerges
from the laser field. These electrons were called back rescattered
ridge (BRR) electrons in Morishita \emph{et al}.~\cite{toru08}. The
BRR electrons lie on a shifted circle in the photoelectron 2D
momentum spectrum. Let the direction of laser polarization be along
the $z$-axis, and the $y$-axis perpendicular to it. After the
backscattered electron emerges from the laser field, the
photoelectron has momentum components
\begin{eqnarray}
\label{key} k_z=k \cos\theta&=& \pm A_0 \mp k_r \cos\theta_r, \\
\label{key2}k_y=k\sin\theta&=&k_r\sin\theta_r.
\end{eqnarray}
The upper signs in Eq.~(\ref{key}) refer to the right-side ($k_z>0$)
while the lower ones to the left-side ($k_z<0$). For backscattering,
the angle $\theta_r$ is greater than 90$^{\circ}$. These two
equations can be expressed in vector form $\textbf{k}=\pm
A_0\hat{z}+\textbf{k}_r$. This vector relation (the outermost
half-circle) is shown in Fig.~5 where the momentum is measured in
units of $A_0$, and the angles are defined as shown. For $k_z>0$,
the returning electron enters the target from the right. After a
large angle scattering, it is deflected by an angle $\theta_r$. As
the electron exits the laser field, it makes an angle $\theta$ with
respect to the polarization axis.

In Morishita \emph{et al}.~\cite{toru08}, it was argued that if the
rescattering picture is correct, the intensity of electrons along
the BRR should be proportional to the elastic DCS of the target ion
by electrons with incident momentum $k_r$. In~\cite{toru08}, this
model was tested based on the HATI electron spectra calculated from
solving the TDSE, for H, Ne, Ar, and Xe targets. For the rare gas
atoms, each target is represented by a model potential of the form,
Eq.~(9). The same model potential was used to calculate HATI
electron momentum spectra $D(k,\theta)$ and the elastic DCS,
$\sigma(k_r, \theta_r)$. By comparing the normalized yield of
$D(k,\theta)$ along the ridge of $k_r$=1.26 $A_0$ with the DCS,
$\sigma(k_r,\theta_r)$, it was shown that the two indeed agree very
well for the targets tested. The tests have been carried out for
different laser intensities and mean wavelengths. Since the
$D(k,\theta)$ calculated from solving the TDSE are considered
``exact'', the test establishes the validity of attributing HATI
electron momentum spectra to elastic backscattering of the returning
electrons with momentum $k_r$=1.26$A_0$.

\begin{figure}
\mbox{\rotatebox{270}{\myscaleboxa{
\includegraphics{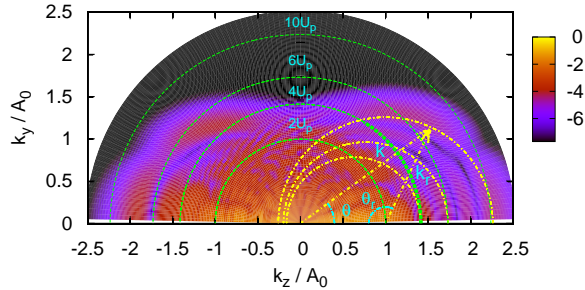}}}}
\caption{(Color online) Typical 2D electron momentum distributions
(in logarithmic scale): The TDSE calculation is for single ionization
of Ar in a 5~fs laser pulse at the intensity of $1.0\times10^{14}$
W/cm$^2$ with the wavelength of 800~nm. Photoelectrons of a given
energy are represented on a concentric circle centered at the
origin. The elastic scattering of a returning electron with momentum
$k_r$ in the laser field is represented by a partial circle with its
center shifted from the origin by $A_r=k_r$/1.26. High-energy
plateau electrons are obtained via large-angle backscattering only.
See text.}
\end{figure}

The theoretical result of \cite{toru08} has been limited to BRR
electrons only where $k_r$=1.26$A_0$, thus it leaves out a large
portion of the HATI spectra where $k_r<1.26A_0$. For these lower
energy electrons, as shown in Section II.A, the electrons may return
to the ion core by following a long- or a short-trajectory. The
ratio of the returning electron momentum $k_r(t_r)$ vs the vector
potential $A_r$=$A(t_r)$ at the time of return, $t=t_r$, are shown
in Fig.~1(b). Note that the ratio does not change significantly in
the time window for those returning electrons  that can be
backscattered to emerge with energies higher than $4U_p$. Thus we
set the relation $k_r=1.26A_r$ for all  HATI electrons, i.e.,
Eqs.~(\ref{key}) and (\ref{key2}) are generalized to
\begin{eqnarray}
\label{qrs1}k_z=k\cos\theta&=& \pm k_r/1.26\mp k_r \cos\theta_r, \\
\label{qrs2}k_y=k\sin\theta&=&k_r\sin \theta_r.
\end{eqnarray}
Recall that if one neglects the effect of core potential, the
returning electron momentum should be determined by the difference
of the vector potentials at the return time and the born time. From
Fig.~1(a), the range of born time is very narrow, thus the same
relation between $k_r$ and $A_r$ for all HATI electrons is a good
approximation. With this model, the center of the circle for each
momentum $k_r$ is shifted by $k_r/1.26$. This has important
implications since the laser parameters such as peak intensity or
wavelength do not enter explicitly in Eq.~(\ref{qrs1}) or
(\ref{qrs2}) any more. How good is this model? We test its validity
using accurate numerical results from TDSE calculations.

\begin{figure}
\mbox{\rotatebox{270}{\myscaleboxa{
\includegraphics{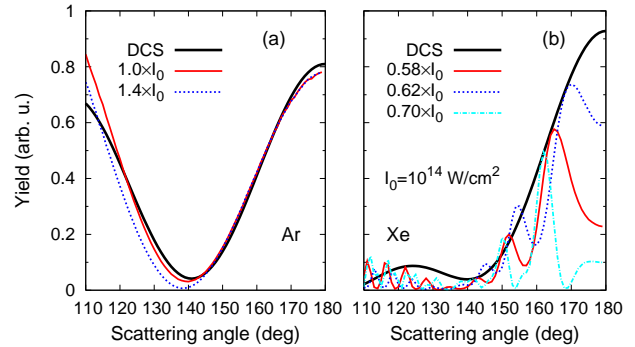}}}}
\caption{(Color online) Electron yield extracted for fixed $k_r$
from the 2D electron momentum distributions for Ar and Xe by solving
the TDSE compared with the corresponding elastic scattering DCS's
(thick black solid lines). (a) For the target of Ar and $k_r=1.22$
by a 5-cycle laser pulse with the wavelength of 800~nm at
intensities of  1.0  and $1.4\times10^{14}$ W/cm$^2$, respectively;
(b) For the target of Xe and $k_r=0.92$ by a 8-cycle laser pulse
with the wavelength of 800~nm at intensities of  5.8 ,  6.2  and
$7.0\times10^{13}$ W/cm$^2$, respectively.}
\end{figure}

Recall that the validity of Eq.~(\ref{QRS}), Eqs.~(\ref{qrs1})
and (\ref{qrs2}), has been fully tested for the case of short pulses
and for $k_r=1.26A_0$ in \cite{toru08}. In Fig.~6(a), we show the
theoretically calculated DCS for Ar at $k_r=1.22$ and compare with
the DCS extracted from HATI momentum spectra obtained from solving
the TDSE for Ar target. In one case we use a five-cycle pulse with
800~nm mean wavelength, and peak intensity of $1.0\times10^{14}$
W/cm$^2$. This is the same as the BRR discussed in \cite{toru08}. In
another case, the peak intensity used is $1.4\times10^{14}$
W/cm$^2$. The BRR electron momentum for the latter is 1.32. If we
extract the DCS from the HATI spectra of the latter at $k_r$=1.22,
as seen from Fig.~6(a), the results are still quite good.

Up to now, we have focused on short laser pulses. For longer pulses,
the electrons generated from different optical cycles of the laser
can interfere. Such interference would result in the well-known ATI
peaks which are separated by the photon energy of the laser. From
Fig.~5, along $k_r$=constant, the photoelectron energy changes as
$\theta_r$ is varied. Thus the assumption that $W(k_r)$ is constant
along a fixed $k_r$ is no longer correct because of the interference
in the wave packet. However, the interference effect is well
behaved. In Fig.~6(b), we show the DCS for e-Xe$^+$ at $k_r$=0.92.
We also extracted the DCS at the same $k_r$=0.92 using the HATI
spectra generated by three lasers, with peak intensity of 7.0, 6.2,
and 5.8 in units of $1.0\times10^{13}$ W/cm$^2$, for 800 nm lasers
of durations of 8 cycles. By assuming a constant wave packet
$W(k_r)$, the ``extracted'' DCS oscillates but the peak positions of
the oscillation still follow the DCS quite accurately (after
normalized). Such oscillations appear to be worrisome in attempts to
extract DCS from HATI spectra using longer laser pulses. However, as
will be shown in Section IV.D, this is not a problem for extracting
DCS from experimental HATI spectra since experiments
``intrinsically'' integrate electron spectra generated over a
distribution of laser intensities. More discussions on the
oscillations of the wave packet are given in the next
subsection.

\subsection{C. Extracting rescattering wave packet from the HATI spectra}

Given the relation between $(k,\theta)$ and $(k_r,\theta_r)$ in
Eqs.~(\ref{qrs1}) and (\ref{qrs2}), in general one can write
$D(k,\theta)=W(k_r,\theta_r)\sigma(k_r,\theta_r)$. If the
rescattering concept is meaningful for a given $k_r$, we expect
$W(k_r,\theta_r)=W(k_r)$, i.e., the rescattering wave packet
distribution is independent of scattering angles. To illustrate this
point, we calculate
\begin{eqnarray}
W(k_r,\theta_r)=D(k,\theta)/\sigma(k_r,\theta_r),
\end{eqnarray}
where $D(k,\theta)$ is obtained from   TDSE and $\sigma(k_r,
\theta_r)$ from Eq.~(\ref{dcs}). The results for $W(k_r,\theta_r)$
are shown for the angular range of $\theta=155^\circ$ to 180$^\circ$
in Figs.~7(a) and 7(b), for two intensities (with other laser
parameters given in the figure), respectively. It is quite clear
that there is little angular dependence of $\theta_r$ such that we
can write $W(k_r,\theta_r)=W(k_r)$. We emphasize that this relation
is based on computational results where $D(k,\theta)$ and
$\sigma(k_r,\theta_r)$ are calculated ``exactly''. The correctness
of $W(k_r,\theta_r)=W(k_r)$ justifies the relation in
Eqs.~(\ref{qrs1}) and (\ref{qrs2}) and it provides a strong
statement of the QRS model for HATI electron momentum spectra, as
stated in Eq.~(\ref{QRS}) in the Introduction. Note that $W(k_r)$ is
extracted from $D(k,\theta)$ at the end of the laser pulse, thus
$W(k_r)$ includes all the quantum interference due to the long- and
short-trajectory electrons, and interference due to wave packets
generated from different optical cycles. In Figs.~7(a,b), the pulse
duration is 5~fs. The oscillation  in the wave packet is due to
interference of long- and short-trajectory electrons that return
with the same $k_r$. As the laser intensity increases, the
oscillations  become faster. This increase of oscillations can be
easily understood based on the SFA2. Note that there are two wave
packets, $W(k_r)$, one from the left and the other from the right
toward the target.
\begin{figure}
\mbox{\rotatebox{270}{\myscaleboxa{
\includegraphics{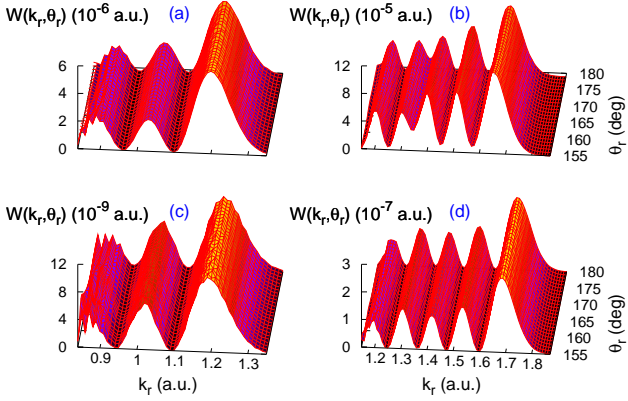}}}}
\caption{(Color online) Right-side wave packets($k_z>0$) extracted
from the TDSE and SFA2 electron momentum distributions for single
ionization of Ar in a 5~fs laser pulse with the wavelength of
800~nm. (a,c) TDSE and SFA2 results at the intensity of
$1.0\times10^{14}$ W/cm$^2$. (b,d) TDSE and SFA2 results at the
intensity of $2.0\times10^{14}$ W/cm$^2$.}
\end{figure}

\begin{figure}
\mbox{\rotatebox{270}{\myscaleboxa{
\includegraphics{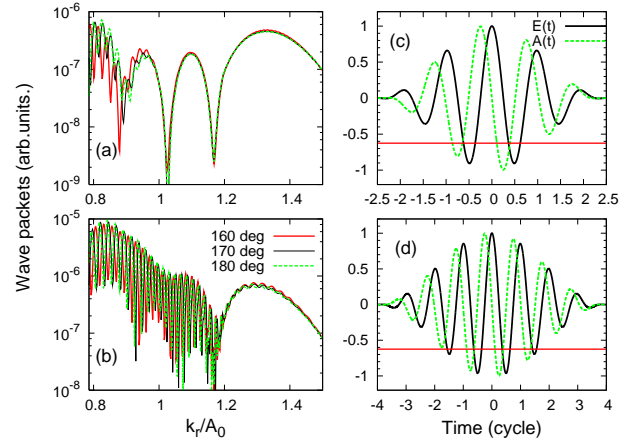}}}}
\caption{(Color online) (a) Right-side wave packets ($k_z>0$)
extracted from TDSE for single ionization of Ar in a 5-cycle pulse
with the wavelength of 800~nm at the intensity of $1.0\times10^{14}$
W/cm$^2$ at $\theta_r$=160$^\circ$, 170$^\circ$ and 180$^\circ$,
respectively; (b) Same as (a) but for an 8-cycle pulse; (c,d)
Electric field and vector potential used in (a,b), respectively. The
peak values of both $E(t)$ and $A(t)$ are normalized to 1. The
horizontal lines in (c,d) are for $A_r$=0.625$A_0$. }
\end{figure}

As discussed earlier, in strong field approximation, SFA2 dominates
over SFA1 for HATI electrons with energies above 4$U_p$, see Fig.~2.
Since rescattering is included in SFA2, we also check whether
separation similar to Eq.~(\ref{QRS}) also applicable to
$D(k,\theta)$ calculated from SFA2. In SFA2 the elastic scattering
of RWP with the target ion is treated to first order only, thus the
corresponding $\sigma(k_r,\theta_r)$ is calculated using the
plane-wave Born approximation, Eq.~(\ref{pwba}). Following the same
procedure as for the TDSE results, we extract the RWP from SFA2. The
results are shown in Figs.~7(c) and 7(d) for the same two sets of
lasers used in the TDSE calculations. Comparing Figs.~7(a,c) and
Figs.~7(b,d), respectively, we note that the RWP's extracted from
SFA2 and from TDSE have very similar shape. After normalization,
their dependence on $k_r$ is nearly identical for the same laser.
The absolute value of $W(k_r)$ from SFA2 is smaller in general since
ionization yield calculated using strong field approximation is
smaller in general. The similarity of $W(k_r)$ from SFA2 and from
TDSE also allows us to interpret the increase of oscillations of RWP
as the laser intensity is increased, as seen in Fig.~7. This
increase can be traced to the actions $S$, Eqs.~(\ref{action1}) and
(\ref{action2}), whose values increase quadratically with the vector
potential, or linearly with the ponderomotive energy $U_p$ and time
duration of the pulse.

For longer pulses, the RWP should reflect the interference from
different optical cycles. Consider the wave packets shown in
Figs.~8(a,b), generated by an 800~nm, peak intensity of
$1.0\times10^{14}$ W/cm$^2$, but one with 5, and the other 8 optical
cycles, respectively. The corresponding $E$-field and $A$-vector are
shown to the right. The horizontal lines in Figs.~8(c,d) are for
$A_r=0.625A_0$. Electrons return with this $A_r$ have $k_r=1.26A_r$
that can be  back rescattered to reach HATI energy of $4U_p$. Note
that $A_r$ determines the photoelectron energy, but the yield is
determined by the electric field at about 3/4 cycles earlier. Thus
for the 5-cycle pulse, HATI electrons are generated from one optical
cycle only, and the oscillation seen in Fig.~8(a) is due to
interference from long- and short-trajectory electrons. For the
8-cycle pulse, at least two optical cycles make contributions to the
HATI spectra, thus interference seen in Fig.~8(b) becomes much more
numerous. The fast oscillations from electrons generated at
different optical cycles overwhelm the slower oscillations from
electrons generated within the same cycle, such that the long- and
short-trajectory interference is seen as the oscillation of the
envelope in $W(k_r)$. As the pulse duration increases, the momentum
distribution of the RWP will become flatter except for the cutoff
region. As demonstrated in Fig.~8(b), the finer oscillations in the
RWP extracted from different angles $\theta_r$ will not be the same.
However, the envelope of the wave packet is independent of
$\theta_r$, such that a single wave packet is still meaningful. More
examples can be seen in \cite{chen_07}.

\subsection{D. Target independence of the shape of the returning electron wave packet}

The fact that the $W(k_r)$ extracted from SFA2 is similar in shape
to that extracted from TDSE is again a consequence of the validity
of the rescattering model. While the absolute returning electron
yield is determined by the initial tunnel ionization rate, its
momentum distribution  $W(k_r)$ is determined almost entirely by the
laser field. In SFA2, this interaction is fully included. The
electron-target ion interaction, which is included in TDSE, affects
$W(k_r)$ weakly only since the returning electron spends most of the
time away from the target ion where the field is dominated by the
laser's electric field.

\begin{figure}
\mbox{\rotatebox{270}{\myscaleboxa{
\includegraphics{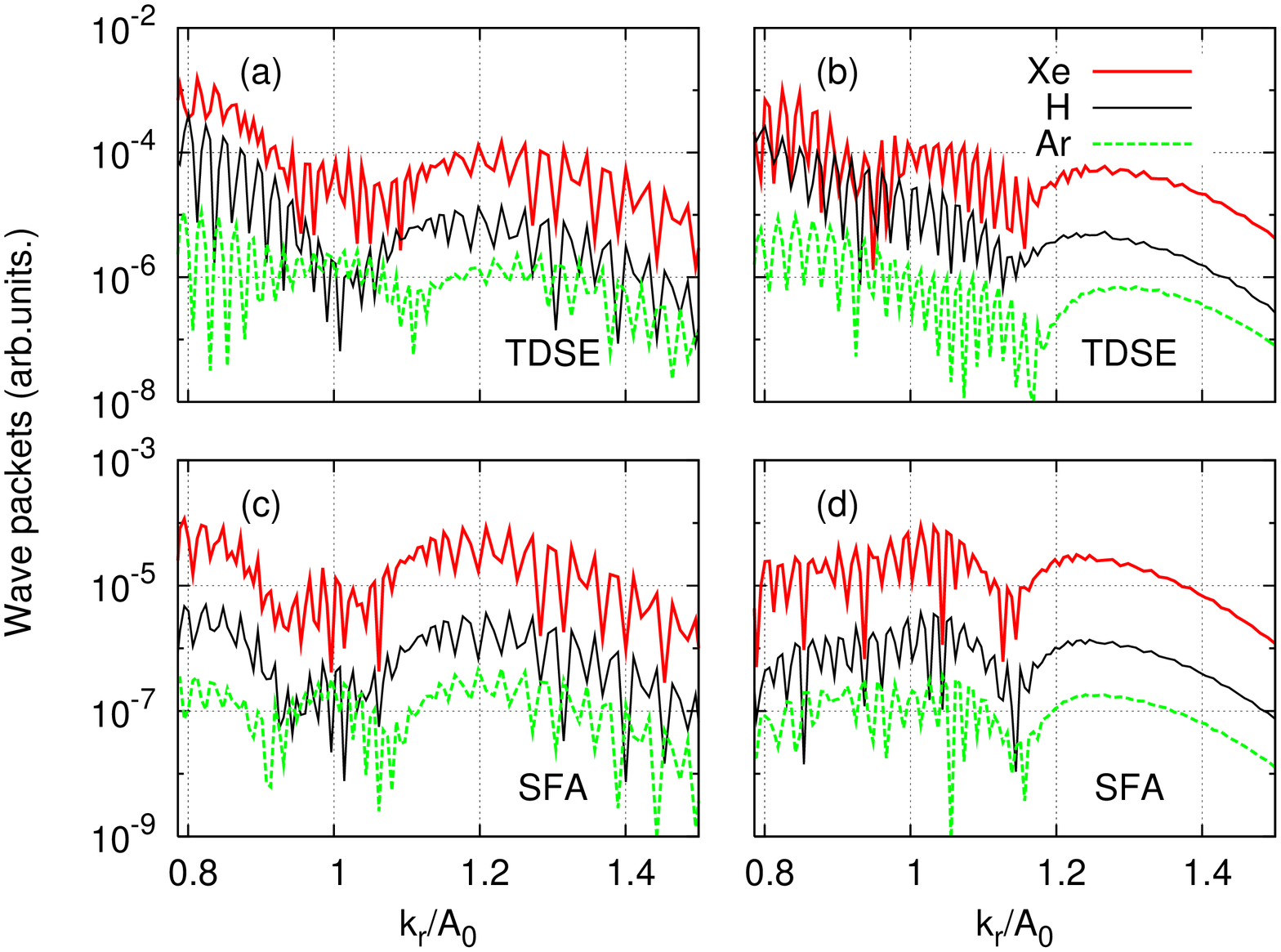}}}}
\mbox{\rotatebox{270}{\myscaleboxa{
\includegraphics{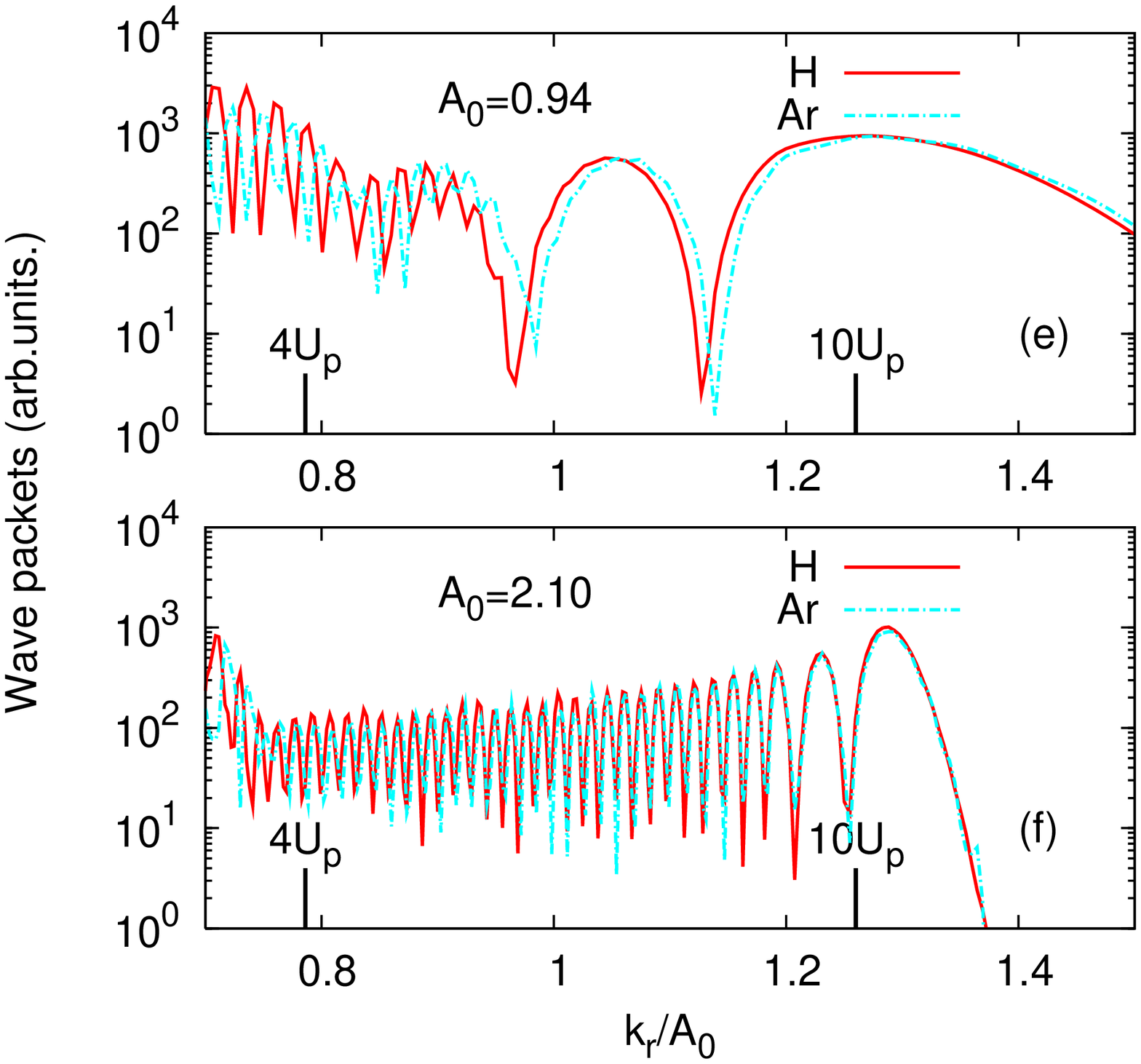}}}}
\caption{(Color online) (a) Left-side wave packets ( $k_z<0$)
extracted from TDSE for single ionization of H, Ar and Xe in a
8-cycle pulse at the peak intensity of $1.0\times10^{14}$ W/cm$^2$
with wavelength of 800~nm, respectively; (b) Same as (a) but for the
right-side ($k_z>0$); (c) Same as (a) but from SFA2; (d) Same as (b)
but from SFA2; (e) Right-side wave packets extracted from SFA2 for
single ionization of H and Ar in a 5-cycle laser pulse at the
intensity of $1.0\times10^{14}$ W/cm$^2$ with the wavelength of
800~nm; (f) same as (e) but for intensity of $2.1\times10^{14}$
W/cm$^2$ and wavelength of 2000~nm.}
\end{figure}

The fact that $W(k_r)$ can be obtained from SFA2 reasonably
accurately has a far-reaching implication. Before discussing such
implications, first we look more carefully at how well the $W(k_r)$
obtained from TDSE and SFA2 agree, for different target atoms. In
Figs.~9(a-d) we show these RWP's generated by an 800nm, 8-cycle,
peak intensity $1.0\times10^{14}$ W/cm$^2$ laser pulse for H, Ar and
Xe targets. We show the ``left'' and ``right'' RWP's which are
different for the short pulse used. To first order, all the wave
packets on the left are similar, and all the wave packets on the
right are similar. A more careful examination reveals that there are
differences. Among the different targets calculated using TDSE, we
note that the distribution of $W(k_r)$ tends to shift to higher
$k_r$ as the ionization energy increases. This is expected since the
returning electron is seeing a more attractive potential from the
ion core, and the effect is bigger for lower energy electrons. Still
the effect is only a few percents. For longer pulses or higher
intensities, see Figs.~9(e,f), the wave packets at lower momenta are
mostly very flat where small shift of individual peaks is not very
important. Due to the small difference of the RWP on the target for
a given laser pulse, the wave packet can be generated using SFA2
from a hydrogenlike target, with the effective charge chosen so that
it gives identical binding energy of the target.

\subsection{E. Elastic electron-ion differential cross sections at large angles}

Elastic and inelastic scattering cross sections between free
electrons and atomic ions have been studied in cross-beam or
merged-beam experiments \cite{brotton}. In such experiments, a
well-collimated electron beam with precisely defined energy is
prepared. For neutral atomic or molecular targets which can be
placed in a gas cell, there have been lots of experimental and
theoretical investigations in the past half a century. These studies
tend to focus on sharp features like Feshbach resonances which often
require a careful treatment of electron correlation effects. For the
HATI spectra, the returning electron is described by a wave packet
which has a broad momentum distribution. Here we tend to focus on
the broader energy and the angular dependence, and neglect the
many-electron effects to first order. Within this model, the
calculation of elastic scattering cross sections is quite simple, as
described in Section II.D.

\begin{figure}
\mbox{\rotatebox{270}{\myscaleboxb{
\includegraphics{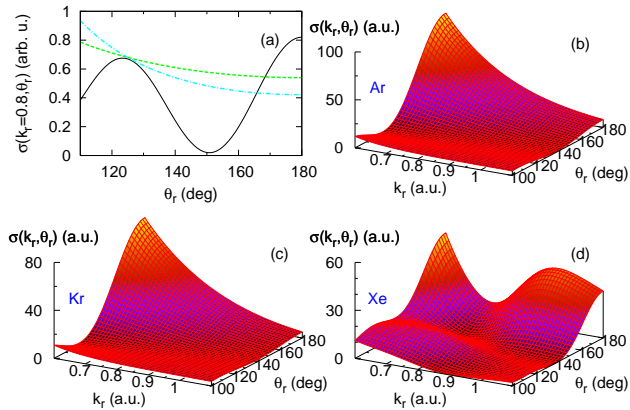}}}}
\caption{(Color online) (a) DCS's for Xe at $k_r=0.8$ in the angular
range of 110$^\circ$-180$^\circ$. Solid line: ``exact'' scattering
wave result [Eq. (42)]; broken line: PWBA [Eq. (45)]; Chain line:
Coulomb wave result [Eq. (41)]. (b,c,d) 2-dimensional ``exact''
scattering wave results of the DCS's for Ar, Kr and Xe,
respectively.}
\end{figure}

In Fig.~10(a) we compare the DCS's calculated for Xe at the incident
momentum of $k_r=0.8$ for scattering angles from
110$^\circ$-180$^\circ$, where the continuum wave functions are
represented by ``exact'' scattering waves, by plane waves and by
Coulomb waves, respectively. The DCS's are normalized near about
130$^\circ$. We note that the DCS calculated from the PWBA is rather
featureless, so is the DCS calculated using Coulomb wavefunctions,
each drops monotonically with increasing scattering angle. On the
other hand, the DCS calculated from the scattering wave shows
complicated pattern which is the well-known Ramsauer-Townsend
electron diffraction. This illustrates that both plane wave and
Coulomb wave are very poor approximation for describing electron-ion
collisions, especially for electrons which undergo large-angle
scattering. For such large deflection angles, the electron has to
penetrate the ion core, thus seeing the short-range part of the
potential. Without scattering waves, the diffraction by the strong
potential $V_{\text{s}}(r)$ is neglected and the HATI spectra cannot
be correctly reproduced.

In Figs.~10(b-d) we show $\sigma(k_r,\theta_r)$ for Ar, Kr and Xe
for $k_r$ from 0.6 to 1.1. The complicated structure, and the
increase of cross sections close to 180$^\circ$, are quite evident
for all three targets. Such structure would not appear if H is used
as the target.  For the large angles discussed here, the minima in
the DCS come  from the interference of contributions from several
partial waves in the scattering  by the short range potential. At
large angles, the Coulomb scattering amplitude [See
Eq.~(\ref{pure_C})] is relatively small. The interference of the two
amplitudes in Eq.~(\ref{amp}) can also produce interference minimum
shown in typical textbook examples \cite{schiff}, but such minimum
occurs at smaller angles. (See the two minima in the DCS for
e$^-$+Ar$^+$ collisions in Fig.~4(c) of \cite{sam-prl} where the
large-angle minimum at 140$^\circ$ was derived from the HATI spectra
and the small-angle minimum at 89$^\circ$ was observed from
electron-Ar$^+$ colliding beam experiment \cite{brotton}.)

The strong dependence of $\sigma(k_r,\theta_r)$ on the target
potential shows that HATI spectra can be obtained accurately only if
electron-ion scattering is treated accurately in the nonlinear
interactions of lasers with atoms and molecules. Thus TDSE
calculations using approximations where the singularity of the
Coulomb core potential is regularized should be handled with care.
Such regularized models have less effect on the total ionization
yield or the electron spectra at low energies, but it will affect
the HATI spectra since these electrons undergo close collisions with
the target core. This also implies that strong field calculations
where continuum electrons are treated at the level of plane waves
(as in SFA2), Coulomb waves (such as using Coulomb Volkov states),
or in the eikonal approximation, will not be adequate for the
description of the HATI spectra.

\subsection{F. The practical quantitative rescattering model and its validity}

Based on the established validity of the QRS model using TDSE and SFA2,
we now propose a practical QRS model for obtaining HATI electron
momentum spectra. We would apply this model to electron energies
above about 4$U_p$ for the total electron energy spectra. For the energy
dependence at each fixed angle, the lower energy limit where this theory applies can be
relaxed, as illustrated in Fig.~3.

In the practical QRS model, the HATI momentum distribution
$D(k,\theta)$ is calculated using Eq.~(\ref{QRS}). We obtain the DCS
using Eq.~(\ref{dcs}) and the returning electron wave packet from
\begin{eqnarray}
\label{getwp}W(k_r)=D_{\text{SFA2}}(k,\theta)/\sigma_{\text{PWBA}}(k_r,\theta_r),
\end{eqnarray}
where $D_{\text{SFA2}}(k,\theta)=|f_2|^2$ and $f_2$ is calculated
from Eq.~(\ref{2nd-order}), and $\sigma_{\text{PWBA}}(k_r,\theta_r)$
is calculated from Eq.~(\ref{pwba}). The relation between
$(k,\theta)$ and $(k_r,\theta_r)$ are given by Eqs.~(\ref{qrs1}) and
(\ref{qrs2}). To obtain $W(k_r)$ from Eq.~(\ref{getwp}), we need to
do SFA2 calculation at one angle $\theta_r$ only. We typically use
$\theta_r$=170$^\circ$.

Since $W(k_r)$ is nearly independent of the target for a given laser
pulse, we can also perform SFA2 calculation using hydrogenic
potential with an effective charge that reproduces the binding
energy of the target atom. We emphasize that using the QRS, the
absolute yield is not obtained. This is also true for most of the
energy and momentum spectra reported in experiments. Using the
practical QRS model, the calculation of HTAI spectra can be a factor
of thousands faster compared to TDSE calculations. However, this is
useful only if we can show that the QRS  reproduces the HATI
momentum and energy spectra at the level comparable to TDSE results.

\begin{figure}
\mbox{\rotatebox{270}{\myscaleboxa{
\includegraphics{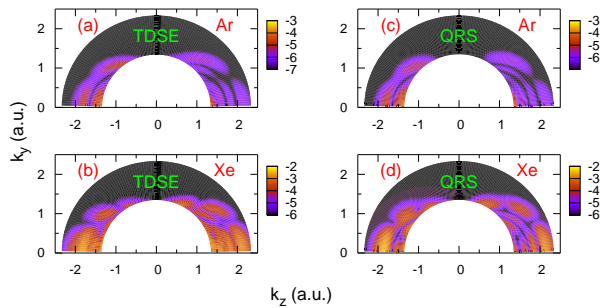}}}}
\caption{(Color online) Logarithmic photoelectron 2D momentum
distributions by a 5-cycle pulse at the peak intensity of
$1.0\times10^{14}$ W/cm$^2$ with wavelength of 800~nm. (a) TDSE
results for Ar; (b) QRS results for Ar; (c) TDSE results for Xe; (d)
QRS results for Xe.}
\end{figure}

In Fig.~11 we show the 2D electron momentum spectra calculated using
QRS and TDSE for Ar and Xe using the laser parameters indicated in
the figure. Recall that for short pulses the left and right wave
packets have to be calculated separately. One can see the overall
agreement between the QRS and TDSE calculations. In making color
plots we renormalize the spectra and the same color schemes are used
in the figures. Since the same laser is used in the calculations for
both Ar and Xe, according to the QRS, the difference in
Figs.~11(c,d) are almost entirely due to the difference in the DCS
[see Figs.~10(b,d)] by the returning electrons colliding with the
target ions of Ar and Xe, respectively.

\begin{figure}
\mbox{\rotatebox{270}{\myscaleboxa{
\includegraphics{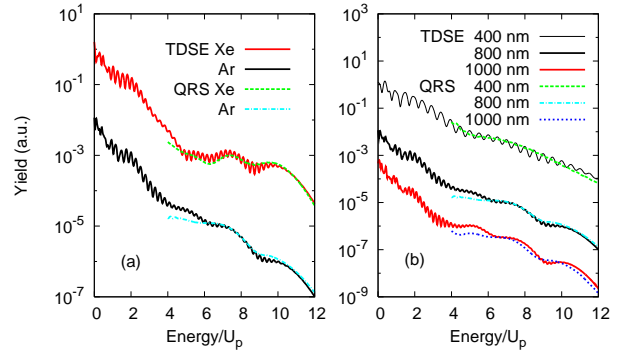}}}}
\caption{(Color online) Comparison of angle-integrated photoelectron
energy spectra from TDSE and QRS. The TDSE spectra are plotted from
0 to $12U_p$ and the QRS results are shown from 4 to $12U_p$. The
QRS spectra are relative and normalized to TDSE results at high
energies. (a) Single ionization of Ar and Xe in a 5-cycle pulse at
the peak intensity of $1.0\times10^{14}$ W/cm$^2$ with wavelength of
800~nm; (b) Single ionization of Ar in a 5-cycle pulse at peak
intensities of 4.0, 1.0 and $0.64\times10^{14}$ W/cm$^2$ with
wavelengths of 400, 800 and 1000~nm, respectively.}
\end{figure}

To display the comparison quantitatively, we show the total electron
energy spectra obtained from TDSE and those from QRS for energies
above $4U_p$. In Fig.~12(a), the electron energy spectra are
obtained by integrating the momentum spectra in Fig.~11. The QRS and
TDSE results agree quite well above $4U_p$. For Ar, we see some
discrepancy close to $4U_p$. In Fig.~12(b), we compare the total
electron spectra using lasers of three different wavelengths, with
the same number of cycles, but with the intensity adjusted such that
$U_p=6$ eV for each pulse, and the Keldysh parameter is 1.14. The
electron spectra calculated from TDSE are placed on the absolute
scale, while the QRS results are normalized individually to achieve
best agreement. Among the three cases, the DCS's are the same. Even
though the returning electron wave packets cover the same range of
momentum, the interference features in $W(k_r)$ are different. Such
interference features are easily reproduced in the wave packet
calculated from SFA2. For 400~nm, the TDSE results show clear ATI
peaks which are not reproduced in the QRS model, as explained in
Section III.D. For additional examples of the comparison between the
QRS model and TDSE, see Chen \emph{et al}.~\cite{chen_08}.

\section{IV. Comparing QRS model with experimental ATI electron spectra}

\subsection{A. Volume effect in experimental ATI spectra}

In the previous Section we compared the HATI spectra calculated
using TDSE and QRS for a given laser pulse with single intensity.
Experimentally, the intensity distribution of a focused laser beam
is not uniform in space. The HATI electrons are collected from the
whole focused volume. Thus to compare with experimental HATI
spectra, theoretical calculations must include the volume effects
\cite{AA26,toru_07}. For a peak intensity $I_0$ at the focal point,
the yield of the photoelectrons with momentum $\textbf{k}$ should be
\begin{eqnarray}
\label{vol}S(\textbf{k},I_0)=\rho\int_{0}^{I_0} D_I(k,\theta)
\left(\frac{\partial V}{\partial I}\right)d I
\end{eqnarray}
where $\rho$ is the density of atoms in the chamber, $D_I(k,\theta)$
denotes the momentum distribution for a single intensity $I$ and
$(\partial V/\partial I)$$d I$ represents the volume of an
isointensity shell between $I$ and $I=I+d I$ defined in \cite{Augst}
for a Lorentzian (propagation direction) and a Gaussian (transverse
direction) beam profile. We use the trapezoidal rule for the
integration over intensity with sufficiently small step size of
$0.01\times 10^{14}$ W/cm$^2$.

In the QRS calculations, we obtain the volume-integrated returning
electron wave packet using Eq.~(\ref{QRS}) since the DCS does not
depend on the laser intensity. Consequently, Eq.~(\ref{vol}) becomes
\begin{eqnarray}
S(\textbf{k},I_0)=\bar{W}_{I_0}(k_r)\sigma(k_r,\theta_r)
\end{eqnarray}
where $\bar{W}_{I_0}(k_r)$ is the volume-integrated wave packet at
the peak intensity $I_0$
\begin{eqnarray}
\bar{W}_{I_0}(k_r)=\rho\int_{0}^{I_0} W_I(k_r) \left(\frac{\partial
V}{\partial I}\right)d I
\end{eqnarray}
with $W_I(k_r)$ being the wave packet for the laser pulse at a
single intensity $I$.

\subsection{B. Wavelength dependence}

In a recent experiment, Colosimo \emph{et al}. \cite{nature08}
reported electron spectra generated from Ar by infrared (IR) to
mid-infrared (MIR) lasers with the same peak intensity, for
wavelength of 800, 1300, 2000 and 3600~nm, respectively. The
measured electron spectra with electron energies in units of $U_p$
are shown in Fig.~14, as well as the results from the QRS model
above $4U_p$ where volume integration effect has been included.
First we note that there is a general agreement between the HATI
spectra from the measurement and from the QRS model. For 800~nm, the
QRS underestimates the 4-$5.5U_p$ region which could be due to the
resonantlike enhancement effect \cite{ang-jpn}. In the figure, the
electron spectra from the different wavelengths are normalized near
at threshold. One notes that the HATI yields decrease rapidly with
increasing wavelength. Such decrease is also familiar in the HHG
spectra \cite{tate}, empirically estimated to decrease like
$\lambda^{-5.5}$. For longer wavelength lasers, the excursion
distance and the return time of the electron after tunnel ionization
both scale with the wavelength, thus increase the effect of
broadening in the returning wave packet.

\begin{figure}
\mbox{\rotatebox{270}{\myscaleboxa{
\includegraphics{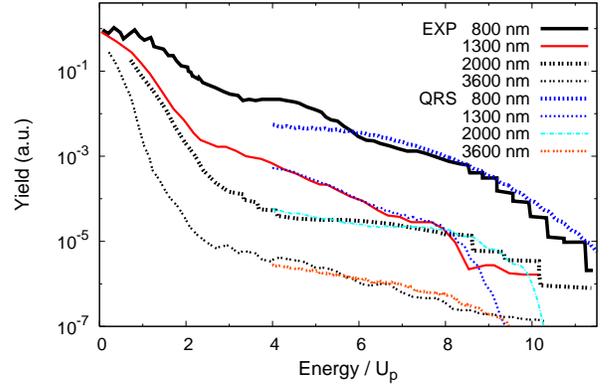}}}}
\caption{(Color online) Comparison of experimental and QRS
angle-integrated photoelectron energy spectra for single ionization
of Ar by few-cycle laser pulses at the peak intensity of
$0.8\times10^{14}$ W/cm$^2$ with wavelengths of 800~nm, 1300~nm,
2000~nm and 3600~nm, respectively. The experimental measurements are
taken from \cite{nature08}. The QRS results starting from $4U_p$ are
normalized to the experimental data individually at high energies to
get best fit. Electron energies are expressed in units of $U_p$.}
\end{figure}

\begin{figure}
\mbox{\rotatebox{270}{\myscaleboxa{
\includegraphics{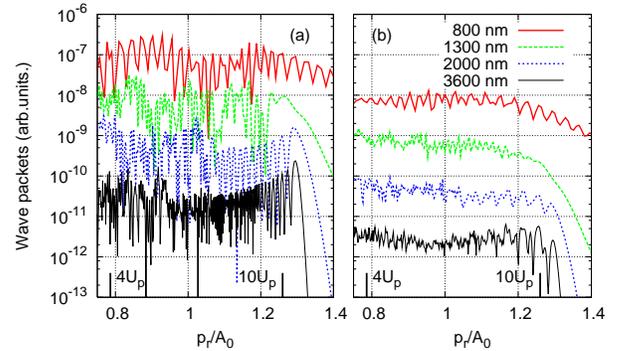}}}}
\caption{(Color online) Wave packets at 800~nm, 1300~nm, 2000~nm and
3600~nm. (a) For single intensity of $0.8\times10^{14}$ W/cm$^2$;
(b) Volume-integrated for the peak intensity of $0.8\times10^{14}$
W/cm$^2$ at the focal point of a Gaussian beam. See text.}
\end{figure}

\begin{figure}
\mbox{\rotatebox{270}{\myscaleboxb{
\includegraphics{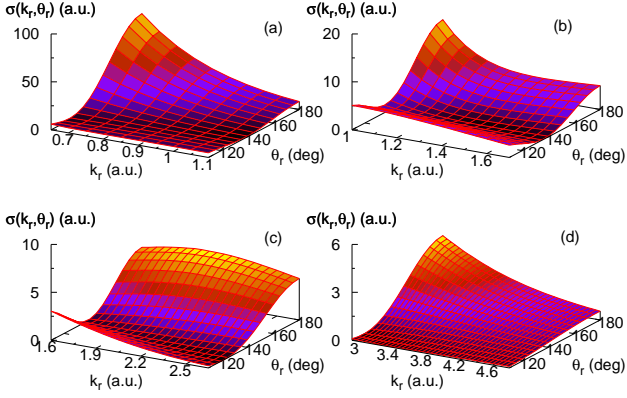}}}}
\caption{(Color online) DCS's for elastic scattering of electron
with Ar$^+$ at large scattering angles of 110$^\circ$-180$^\circ$.
The incident electron momenta are: (a) 0.65-1.10, (b) 1.00-1.75, (c)
1.60-2.65, and (d) 2.95-4.80, corresponding to the high energy
plateau regions for 800~nm, 1300~nm, 2000~nm and 3600~nm,
respectively.}
\end{figure}

From Fig.~13, we note that the slope of the electron spectra appears
to flatten out considerably in going from 1300 nm to 2000 nm. To
understand the origin of the slope change, we show in Fig.~14(a) the
returning electron wave packet obtained from the QRS model. The
momentum of the wave packet is expressed in units of $A_0$ of the
laser pulse at the focus center. We note the rapid oscillations are
due to the ATI peaks. In Fig.~14(b) the volume integrated wave
packets are shown. We note that they are mostly flat above $4U_p$.
According to the QRS model, Eq.~(53), we thus expect that the slope
change in the experimental data is due to the elastic scattering
cross sections. In Fig.~15, we show the differential cross sections
of electron-Ar$^+$ collisions in the four momentum ranges for the
returning electrons that contribute to the HATI spectra for the four
wavelengths used. Note that the DCS peaks sharply at large angles
near 180$^\circ$ for the momentum range of 1.60-2.65 for the 2000~nm
laser pulse. Such behavior of the DCS is responsible for the much
flatter HATI spectra seen in Fig.~15 for the 2000 nm pulse. Note the
similarity of the DCS in Fig.~15(c) in Ar, and the DCS in Xe in
Fig.~12(a). The large DCS at angles close to 180$^\circ$ is
responsible for the much flatter energy dependence observed in the
HATI spectra.

The above analysis shows that the QRS model not only can explain
experimental results, but also offers a clear physical
interpretation of the origin in the difference of the observed HATI
spectra. We comment that in principle these HATI spectra can also be
calculated by solving the TDSE. However, as shown in Colosimo
\emph{et al}. \cite{nature08}, such TDSE calculations for MIR lasers
are very difficult due to their large long excursion distance and
the need of a large box in the calculation for MIR laser pulses.
Using the QRS, the calculation is much easier. In fact, the volume
integrated returning wave packet is quite flat that one may even
just approximate it by a constant. The QRS model then would predict
that the slope seen in Fig.~13 is due to the integration of the DCS
over the scattering angles. Note that by changing the peak intensity
at the laser focus, the same flat plateau seen in Fig.~13 is
expected to shift to different wavelength.

\subsection{C. Target and intensity dependence}

We also used the QRS model to simulate HATI spectra from some
earlier experiments of Grasbon \emph{et al}. \cite{Gra03}. Since in
experiments the peak laser intensity is often determined
approximately only, we perform the simulation by treating the peak
laser intensity as a free parameter. We assume that all the
electrons from the focal volume are collected in the experiment. We
use the pulse duration and wavelength reported in the experiment. In
Fig.~16(a) we compare the experimental HATI spectra measured with
the ones obtained from the QRS model. Note that the peak intensity
reported in the experiment is often different from the one that
gives the best fit obtained from the QRS. In each spectra, we
normalized the QRS result such that it gives best overall fit to the
experimental data. Fig.~16(b) illustrates the example how the slope
of the HATI spectra changes as the peak intensity at the focus is
varied.

\begin{figure}
\mbox{\rotatebox{270}{\myscaleboxa{
\includegraphics{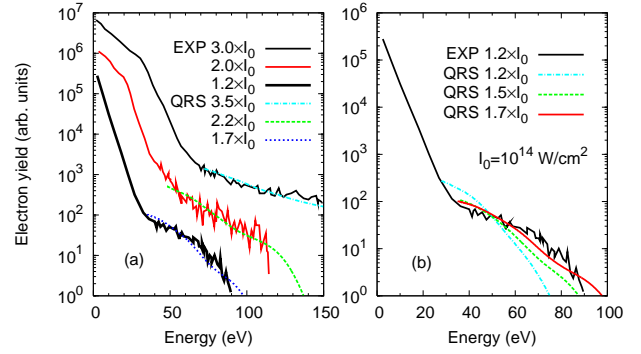}}}}
\caption{(Color online) (a) Comparison of the total energy spectra
of QRS with experimental measurements for Ar of \cite{Gra03} at
different intensities; (b) Illustration of how to adjust intensity
in the QRS simulation to fit experimental measurement. In the QRS
calculations, volume effect has been included. }
\end{figure}

\begin{figure}
\mbox{\rotatebox{270}{\myscaleboxa{
\includegraphics{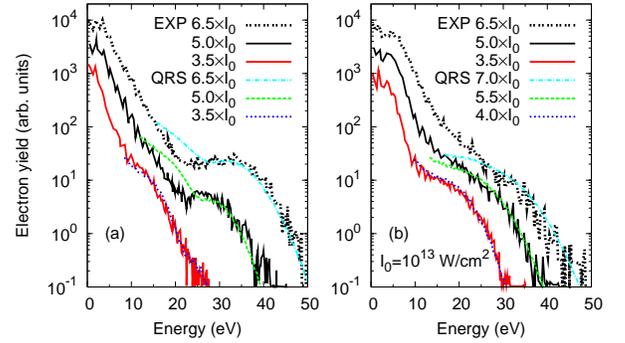}}}}
\caption{(Color online) Comparison of the total electron energy
spectra from QRS with the experimental measurements of \cite{Gra03}.
(a) Xe; (b) Kr. In the QRS calculations, volume effect has been
included.}
\end{figure}

Similar analysis has been carried out for Xe and Kr targets, with
the results shown in Fig.~17. In Kr, the agreement is very good for
all three spectra. The small difference at the low energy end could
be due to contributions from  direct tunneling ionization. For Xe,
we have been able to reproduce the outer part of the plateau well.
At lower electron energies the experimental data show a deeper
minimum for the two upper intensities which are not reproduced by
the QRS simulation. The reason for this discrepancy is not clear at
this time. In Fig.~18(a) we show the relative weights of the
different peak intensities that contribute to the total ionization
yields for the focal peak intensity of $0.65\times10^{14}$ W/cm$^2$.
In Fig.~18(b), we show that the HATI spectra from the QRS agree well
with the result from solving the TDSE at peak intensity of
$0.59\times10^{14}$ W/cm$^2$, the intensity that contributes most to
the electron yields, see Fig.~18(a).

\begin{figure}
\mbox{\rotatebox{270}{\myscaleboxa{
\includegraphics{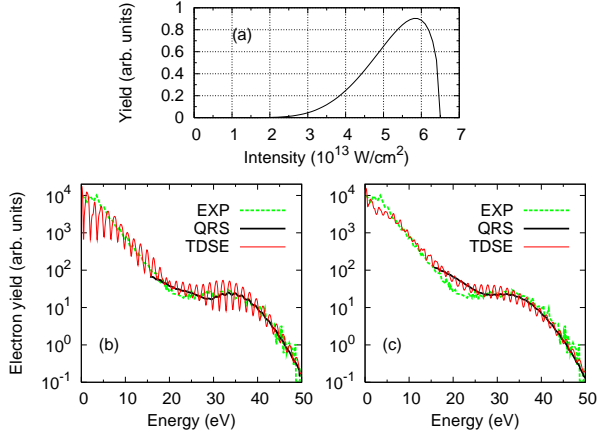}}}}
\caption{(Color online) Single ionization of Xe by a 8~fs pulse at
the peak intensity of $0.65\times10^{14}$ W/cm$^2$ with wavelength
of 800~nm. (a) Laser volume effect analysis showing the relative
contributions of laser peak intensity to the generated electron
spectra.  (b) Comparison of QRS with TDSE at the single intensity of
$0.59\times10^{14}$ W/cm$^2$; (c) Comparison of QRS with TDSE with
volume effect included. }
\end{figure}

Fig.~18(c) shows that volume integrated electron spectra calculated
from TDSE and from the QRS agree well, but both cannot reproduce the
deeper experimental minimum. This minimum occurs at about 20~eV,
which is close to returning electron momentum of $k_r=0.67$, or
electron energy of about 6~eV. At such a low energy, electron-Xe$^+$
elastic scattering cross sections may not be well described by the
model potential approach. Recall that in QRS and TDSE, we treated Xe
atom using a single active electron model. For electron-atom and
electron-ion collisions, it is generally known that many-electron
correlation effect becomes more important as the collision energy
decreases. (An example where the single electron approximation fails
has been noted in the photodetachment of $H^-$ by MIR lasers, see
Zhou \emph{et al} \cite{zhou_08}.) Thus one possible explanation for
the failure of the QRS theory to reproduce the experimental
observation in Fig.~17(a) is the need to include electron
correlation effect in calculating e$^-$-Xe$^+$ elastic scattering
cross sections. While such calculations have been carried out using
many-body perturbation theory \cite{johnson,johnson2} in a number of
cases, no such results have been reported for the present system.
The fact that the minimum occurs at the same photoelectron energy in
the experimental data at the two upper intensities, see Fig.~17(a),
also offer a hint that the discrepancy is due to error in the DCS
used. The minimum does not appear at the lowest intensity in
Fig.~17(a) since the HATI yield drops rapidly in the same energy
region.

\subsection{D. Extracting DCS from experimental electron spectra and Other Applications}

We have applied the QRS model to a number of other topics involving
HATI electrons so far. In Chen \emph{et al}.~\cite{chen_08} the
flatness of the HATI spectra vs electron energies observed for rare
gas atoms and alkali atoms were investigated and interpreted in
terms of the energy and angular dependence of the DCS, similar to
the examples presented here. The QRS model has also been applied to
retrieve the absolute value of the carrier-envelope-phase of
few-cycle pulses, as well as the pulse duration and the peak laser
intensity in Micheau \emph{et al}. \cite{sam-prl,sam-jpb}. Using the
wave packet extracted from the HATI spectra and electron-impact
ionization cross sections, the nonsequential double ionization of Ar
has been obtained in Micheau \emph{et al}. \cite{sam-NSDI}.

\begin{figure}
\mbox{\rotatebox{270}{\myscaleboxa{
\includegraphics{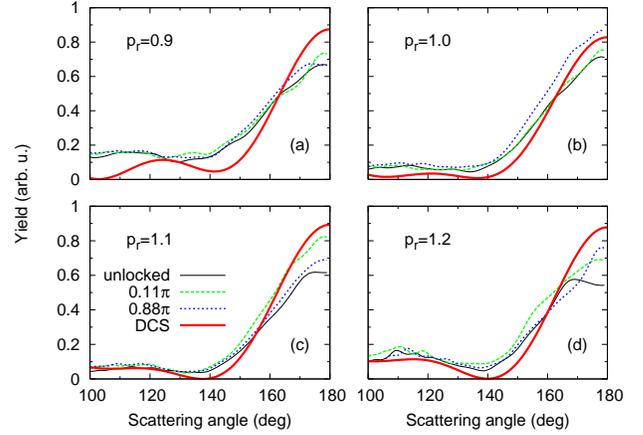}}}}
\caption{(Color online) Comparison of the DCS's with electron yield
extracted from the experimental measurements for Xe by laser pulses
with different CEP's and unlocked CEP for (a) $k_r$=0.9, (b)
$k_r$=1.0, (c) $k_r$=1.1, and (d) $k_r$=1.2. The CEP's are given in
the figure for phase-locked pulses.}
\end{figure}

Following the earlier theoretical paper of Morishita \emph{et
al}.~\cite{toru08} in which DCS was extracted from the HATI spectra
along the BRR, i.e., along $k_r=1.26A_0$, the prediction was
confirmed in two experiments \cite{ueda,ray}. In these experiments,
pulse durations of about 100~fs and 8~fs were used, respectively. As
shown in Fig.~6(b), for such long pulses the extraction of DCS is
difficult because of the interference in the wave packet.
Interestingly, since the experimental electron spectra are collected
from the whole focal volume, the oscillation in the
volume-integrated wave packet at a given $k_r$ is smoothed out. Thus
in effect, the success of retrieving the DCS in these two earlier
experiments \cite{ueda,ray} is based on the present QRS model,
rather than the early theory of \cite{toru08}. In other words, the
effect of volume integration in the electron spectra actually
simplifies the retrieval of the DCS. Similarly, the DCS can also be
extracted from ATI spectra generated by few cycle pulses. For these
pulses, the returning electron wave packet depends on the
carrier-envelope phase. One can obtain the DCS from experiments
where the CEP is locked, or from measurements where the CEP is not
locked.  The extracted DCS should be the same based on the QRS
model. In Fig.~19, we show the extracted DCS for Xe at four
different electron momentum values, using the experimental data of
Kling \emph{et al}. \cite{njp}. The data extracted from these
different data sets agree quite well, and they agree reasonably well
with the theoretical DCS calculated from single active electron
approximation. The theory shows slightly deeper minimum, but the
angular resolution was not considered in the theoretical
calculations. We emphasize that the extracted DCS should be
independent of the lasers used.

\section{Summary and Outlook}

In this paper we present a comprehensive QRS theory for describing
the energy and momentum distributions of HATI electron spectra
generated by intense laser pulses. Although HATI spectra have been
interpreted in terms of rescattering concept since the 1990's, the
QRS model is the first quantitative rescattering theory that can
achieve accuracy comparable to those obtained from solving the TDSE.
The essential ingredient of the QRS is governed by Eq.~(1) which
states that HATI electron momentum distributions can be expressed as
the product of a returning electron wave packet $W(k_r)$ and the
elastic differential cross sections $\sigma(k_r, \theta_r)$ between
\emph{free} electrons and the target ion. The validity of the QRS
model is carefully tested against results obtained from solving TDSE
using atomic targets in the single electron approximation. Here are
a number of the most notable results:

(i) The wave packet $W(k_r)$ is mostly independent of the target,
and can be calculated from the 2nd-order strong field approximation
theory. All the laser dependence in the HATI spectra is governed by
$W(k_r)$. The target structure enters through the DCS, $\sigma(k_r,
\theta_r)$.

(ii) Using a practical QRS model, HATI spectra can be calculated
with accuracy comparable to those obtained from solving TDSE, but
with saving of computing time by a factor of thousands. This is
particularly important when one wants to compare calculations with
experimental measurements where repeated calculations have to be
done for many intensities. With the QRS, as we demonstrated in
Section IV, quantitative comparison of HATI spectra between theory
and experiment is now possible. The QRS theory also is capable of
checking consistency of different experiments.

(iii) The QRS model allows the extraction of the returning electron
wave packet (in the laser field) based on HATI spectra at the end of
the laser pulse, instead of using classical simulations or
theoretical models when the electron is still in the laser field.
Our extracted $W(k_r)$ reflects the interference due to returning
electrons with long- vs short-trajectories, as well as interference
from electrons released from different optical cycles.

(iv) The QRS model allows one to separate the role of laser pulses
which is contained in $W(k_r)$, and the role of the target which is
contained in $\sigma(k_r,\theta_r)$. By calculating $W(k_r)$ using
SFA2, the difficult part of the nonlinear laser-atom interaction is
avoided. This accounts for the major saving in computer time in the
QRS as compared to TDSE. The conceptual separation of the nonlinear
effect of lasers on the wave packet and the structural information
contained in the DCS has many ramifications. In particular, it
should be possible to generalize the QRS model to include
many-electron effects, just by replacing $\sigma(k_r,\theta_r)$ with
the DCS calculated including electron correlation effect. But most
importantly, the QRS model offers an opportunity to calculate
accurate HATI spectra for molecules.

For molecular targets, there are many possible important
applications. One can use a pump pulse to impulsively align
molecules \cite{rmp-align}. By taking the HATI spectra at the time
when the molecules are preferentially aligned or anti-aligned with
respect to the polarization of the pump beam, the dependence of the
DCS on the alignment of molecules can be obtained. Such measurements
are beginning to emerge \cite{meckel}. However, current
interpretation of such experimental data, for molecules that are
aligned \cite{meckel} or not aligned \cite{ueda-jpb}, rely on
intuitive simple physical model \cite{meckel}, or on the extension
of SFA2 to molecular targets \cite{milo-MO-prl,milo-MO-pra}. As
shown in the present paper, SFA2 is not expected to adequately
describe the backscattering of the returning electrons by the
molecular ions. Using the QRS, more accurate description of the HATI
spectra is possible if the DCS from electron-molecular ions
collisions are available. Electron-molecule as well as
electron-molecular ion collisions have been studied over the past
few decades, both in experiments and in theory. A few general
purpose codes are in existence
\cite{lucchese1,lucchese2,tonzani-cpc,tonzani-chris}. By adopting
these codes to obtain the DCS needed for the QRS model, accurate
HATI spectra from aligned molecules can be calculated.

Another potentially very important application is to use infrared
lasers for dynamic imaging of a transient molecule \cite{toru-njp}.
Using a pump pulse to initiate a chemical reaction, the time
evolution of the transient molecule, including the position of its
constituent atoms, can be probed by measuring the HATI spectra by a
probe laser. Today few-cycle pulses of duration of a few
femtoseconds are readily available, and as indicated in Section IV,
no CEP stabilization is needed in order to extract the DCS. From the
dependence of the DCS with respect to the time delay, one may be
able to extract the time evolution of the structure of the transient
molecule, thus achieving dynamic chemical imaging of molecules with
temporal resolution of a few femtoseconds. The feasibility of
extracting the structure of the target from the DCS has already been
demonstrated for rare gas atoms using a simple genetic algorithm
\cite{XU09}. Extension of the method to molecular targets is likely
to be straightforward. Thus while the idea of using laser induced
electron diffraction for imaging the structure of molecules had been
proposed since 1996 \cite{zuo}, due to the relatively low energies
of the returning electrons, the standard electron diffraction theory
\cite{diffra-book} is not applicable. On the other hand, with the
QRS proposed here, and with the implementation of the
state-of-the-art electron-molecule scattering codes, we believe that
the fundamental theoretical tools needed for retrieving the
structure of transient molecules based on laser-induced high-energy
photoelectron spectra have been established.

Before closing, perhaps it is fitting to mention that an expression
similar to Eq.~(1) has been used to explain high-energy electron
emission in the forward directions in energetic ion-atom collisions,
such as the collision between Cu$^{5+}$ with H$_2$ in Liao \emph{et
al}. \cite{liao}. These electrons are best understood in the
projectile frame where a beam of ``free'' electrons from H$_2$ are
incident on the Cu$^{5+}$ ion. When these electrons are
backscattered by the ion, they emerge as high-energy electrons in
the laboratory frame. In this model, the beam of ``free'' electrons
from H$_2$ is treated as a wave packet, represented by the Compton
profile of the target, in the projectile frame, similar to the
returning wave packet $W(k_r)$ in the present QRS model. The
transformation of momentum from the projectile frame to the
laboratory frame in ion-atom collision is equivalent to the $-A_r$
term in the present paper, see Eqs.~(\ref{qrs1}) and (\ref{qrs2}).
In fact the similarity is so close that we quote the abstract of
\cite{liao} here: ``We present a method of deriving energy and
angle-dependent electron-ion elastic scattering cross sections from
doubly differential cross sections for electron emission in ion-atom
collisions. By analyzing the laboratory frame binary encounter
electron production cross sections in energetic ion-atom collisions,
we derive projectile frame differential cross sections for electrons
elastically scattered from highly charged projectile ions in the
range between 60$^{\circ}$ and 180$^{\circ}$. The elastic scattering
cross sections are observed to deviate strongly from the Rutherford
cross sections for electron scattering from bare nuclei. They
exhibit strong Ramsauer-Townsend electron diffraction in the angular
distribution of elastically scattered electrons, providing evidence
for the strong role of screening played in the collision.'' Indeed,
electron scattering experiments can be carried out without directly
preparing a well-collimated electron beam!

\section{Acknowledgment}

This work was supported in part by Chemical Sciences, Geosciences
and Biosciences Division, Office of Basic Energy Sciences, Office of
Science, US Department of Energy. T. M. is also supported by a
Grant-In-Aid for Scientific Research (C) from the Ministry of
Education, Culture, Sports, Science and Technology, Japan, and by
the Japanese Society for the Promotion of Science (JSPS) Bilateral
joint program between the U.S. and Japan.

\end{document}